\documentclass{aa}
\usepackage{psfig}

\newcommand\eg{{\it e.g.} }
\newcommand\etal{et~al.}
\newcommand\ie{{\it i.e.~}\ }

\newcommand\Lya{Ly$\alpha$}
\font\aipsfont = cmsy9 scaled\magstep1

\newcommand\aips {{\aipsfont AIPS}}
\newcommand\Lh {{\aipsfont L$\;$}}
\newcommand\minpoint{$^{\prime}\mskip-4.7mu.\mskip0.8mu$}
\newcommand\araa{{ARA\&A}}
\newcommand\aap{{A\&A}}
\newcommand\aasup{{A\&AS}}
\newcommand\aaps{{A\&AS}}
\newcommand\aj{{AJ}}
\newcommand\apj{{ApJ}}
\newcommand\apjl{{ApJ}}
\newcommand\apjs{{ApJS}}
\newcommand\iaucirc{{IAU Circ.}}
\newcommand\mnras{{MNRAS}}

\newcommand\nat{{Nature}}
\newcommand\pasp{{PASP}}
\def\spose#1{\hbox to 0pt{#1\hss}}
\newcommand\simlt{\mathrel{\spose{\lower 3pt\hbox{$\mathchar"218$}}
     \raise 2.0pt\hbox{$\mathchar"13C$}}}
\newcommand\lesssim{\mathrel{\spose{\lower 3pt\hbox{$\mathchar"218$}}
     \raise 2.0pt\hbox{$\mathchar"13C$}}}
\newcommand\simgt{\mathrel{\spose{\lower 3pt\hbox{$\mathchar"218$}}
     \raise 2.0pt\hbox{$\mathchar"13E$}}}
\newcommand\gtrsim{\mathrel{\spose{\lower 3pt\hbox{$\mathchar"218$}}
     \raise 2.0pt\hbox{$\mathchar"13E$}}}
\newcommand\arcdeg{\degr}
\newcommand\nodata{...}
\begin{document}

\thesaurus{04.19.1; 11.01.2; 13.18.1}

\title{A Sample of 669 Ultra Steep Spectrum Radio Sources to Find High Redshift Radio Galaxies \thanks{Tables A.1, A.2 and A.3 are also and appendices B, C and D are only available in electronic form at the CDS via anonymous ftp to cdsarc.u-strasbg.fr (130.79.128.5) or via http://cdsweb.u-strasbg.fr/Abstract.html }}
\subtitle{}
\titlerunning{A sample of 669 USS sources}

\author{Carlos De Breuck \inst{1,2} \and Wil van Breugel \inst{2} \and Huub J. A. R\"ottgering \inst{1} \and George Miley \inst{1}}

\authorrunning{Carlos De Breuck \etal}

\offprints{Carlos De Breuck}

\institute{Sterrewacht Leiden, Postbus 9513, 2300 RA Leiden, The
             Netherlands; debreuck,miley,rottgeri@strw.leidenuniv.nl
             \and Institute of Geophysics and Planetary Physics,
             Lawrence Livermore National Laboratory, L-413, Livermore,
             CA 94550, U.S.A.; wil@igpp.llnl.gov}

\date{Received 1999 November 5; accepted 2000 February 10}

\maketitle

\begin{abstract}
Since radio sources with Ultra Steep Spectra (USS; $\alpha \lesssim -1.30; S \propto \nu^{\alpha}$) are efficient tracers of high redshift radio galaxies (HzRGs), we have defined three samples of such USS sources using the recently completed WENSS, TEXAS, MRC, NVSS and PMN radio-surveys. Our combined sample contains 669 sources with $S_{1400} > 10$~mJy and covers virtually the entire sky outside the Galactic plane ($|b|>15$\arcdeg). For our 2 largest samples, covering $\delta > -35\arcdeg$, we selected only sources with angular sizes $\Theta < 1\arcmin$. For 410 sources, we present radio-maps with 0\farcs3 to $\sim$5\arcsec\ resolution from VLA and ATCA observations or from the FIRST survey, which allows the optical identification of these radio sources.

Comparison with spectrally unbiased samples at similar flux density levels, shows that our spectral index, flux density, and angular size selections do not affect the angular size distribution of the sample, but do avoid significant contributions by faint foreground spiral galaxies.
We find that the spectral index distribution of 143,000 sources from the WENSS and NVSS consists of a steep spectrum galaxy and a flat spectrum quasar population, with the relative contribution of flat spectrum sources doubling from $S_{1400}>0.1$~Jy to $S_{1400}>2.5$~Jy.
The identification fraction of our USS sources on the POSS ($R \lesssim 20$) is as low as 15\%, independent of spectral index $\alpha < -1.30$. We further show that 85\% of the USS sources that can be identified with an X-ray source are probably contained in galaxy clusters, and that $\alpha < -1.6$ sources are excellent Galactic pulsar candidates, because the percentage of these sources is four times higher in the Galactic plane.

Our sample has been constructed to start an intensive campaign to obtain a large sample of high redshift objects ($z>3$) that is selected in a way that does not suffer from dust extinction or any other optical bias.


\end{abstract}
   
\section{Introduction}
Radio galaxies have now been found out to redshifts of $z=5.19$ (\cite{wvb99b}) and radio-loud quasars out to $z=4.72$ (\cite{hoo98}). Although new optical selection techniques such as color-dropouts, deep spectroscopy of blank fields, and narrow-band \Lya\ imaging have now found galaxies at similar (\cite{ste99}) and even higher redshifts (up to $z \sim 5.75$; \cite{dey98}; \cite{wey98}; \cite{spi98}, \cite{hu99}), radio sources are still the only objects that can be selected uniformly over all redshift ranges, and in a way that does not suffer from optical biases such as dust extinction, which is known to be important at these high redshifts (\eg\ \cite{hug98}; \cite{ivi98}; \cite{dic98}).

At low to moderate redshift ($z \lesssim 1$), powerful radio sources are uniquely identified with massive ellipticals (\cite{lil84}; \cite{owe89}; \cite{bes98}; \cite{mcl00}). The strongest indications that this is also true at higher redshifts comes from the near-IR Hubble $K-z$ diagram of radio galaxies which shows a remarkably close correlation from the present out to $z=5.19$ (\cite{lil89}; \cite{eal97}; \cite{wvb98}, \cite{wvb99b}). This suggests that we can use radio galaxies to study the formation and evolution of the most massive galaxies, which, by their implied star-formation history, can put important constraints on galaxy formation models, and even on cosmological parameters (\eg \cite{dun96}; \cite{spi97}).
Although the unification model for radio galaxies and quasars (\eg \cite{bar89}) suggests we could also use quasars as tracers, a detailed stellar population study of quasar host galaxies is almost impossible due to the extreme luminosity of the AGN. Furthermore, samples of radio sources designed to find large quantities of quasars require additional optical selections (\eg\ \cite{gre96}, \cite{hoo98}, \cite{whi00}).

\begin{figure}
\psfig{file=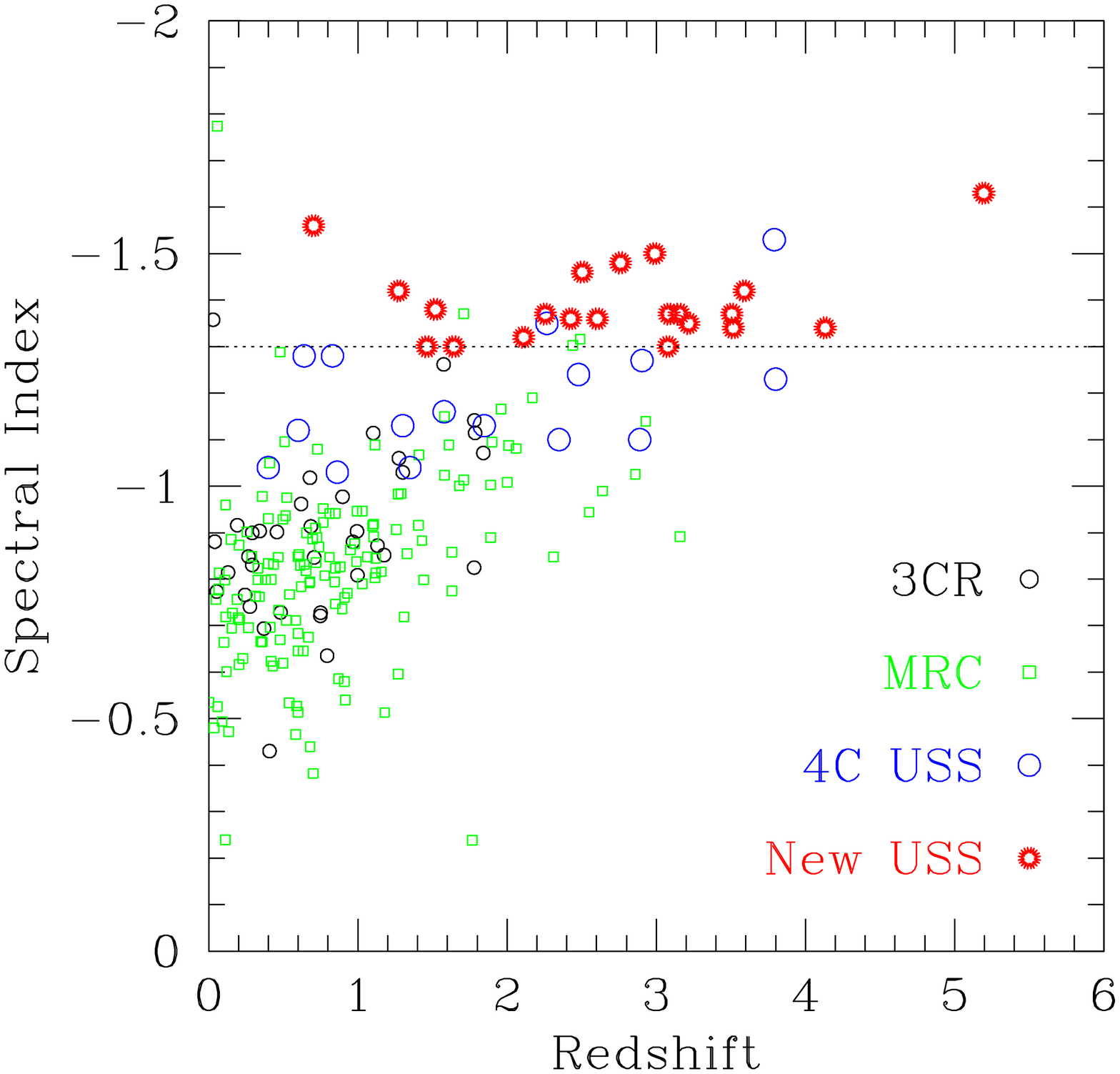,width=8.8cm}
\caption[zalpha.ps]{$\alpha_{1400}^{325}$ against $z$ for 2 samples
without spectral index selection (3CR, \cite{spi85} and MRC,
\cite{mcc96}), and 2 USS samples (4C, \cite{cha96a} and our new WN/TN
samples, as defined in this paper). Note that the correlation is
present in the spectrally unbiased 3CR and MRC, and that the 4C and
our new USS samples are finding three to five times more $z>2$ radio
galaxies than the MRC. The horizontal dotted line indicates the
$\alpha_{1400}^{325} < -1.3$ cutoff used in our USS
sample. \label{zalpha}}
\end{figure}

Considerable effort has been spent over the last decade to find these high redshift radio galaxies (HzRGs), which has lead to the discovery of more than 140 radio galaxies at redshifts $z>2$ (see \eg \cite{deb98a} for a recent summary). However by $z>3$, their numbers become increasingly sparse, and using flux limited radio surveys such as the 3CR ($S_{178} > 10$~Jy; \cite{lai83}), or the MRC strip ($S_{408} > 0.95$~Jy; \cite{mcc96}), the highest redshift radio galaxy found so far is at $z\sim 3.2$ (Fig. \ref{zalpha}; \cite{raw90}; \cite{mcc96}). This redshift limit arises because radio power is correlated with redshift in bright flux limited samples, and an upper limit exists in the radio luminosity. Lowering the flux limit would not only substantially increase the number of sources in these samples, but at the same time the fraction of luminous very high redshift radio galaxies would decrease (\cite{blu98}, \cite{jar99}). This fractional decrease would arise even if there is no decrease in co-moving space density at $z \sim 2.5$. Such a redshift cutoff has been suggested by Bremer \etal (1998), but recently Jarvis \etal (1999) rule out a break at $z\lesssim 2.5$. To efficiently find large numbers of HzRGs in acceptable observing times, it is therefore necessary to apply additional selection criteria, at the expense of completeness.

By far the most successful selection criterion has been the ultra steep spectrum criterion (\eg \cite{rot94}; \cite{cha96a}; \cite{blu98}). Selecting sources with very steep radio spectra increases dramatically the chance of finding $z>2$ radio galaxies (Fig. \ref{zalpha}). This technique is based on the results of Tielens \etal\ (1979) \nocite{tie79} and Blumenthal \& Miley (1979)\nocite{blu79}, who found that the identification fraction on the POSS ($R \lesssim 20$) decreases with steepening spectral index, consistent with the steeper sources being at higher redshifts. It is now getting clear that this correlation can be explained by a combination of a K-correction of a concave radio spectrum and an increasing spectral curvature with redshift (\cite{kro91}, \cite{car99}; \cite{wvb99a}). To further investigate the $z - \alpha$ correlation, we have calculated  spectral indices using the flux densities from the WENSS (\cite{ren97}) and NVSS (\cite{con98}) catalogs for four different samples: the flux density limited 3CR (\cite{spi85}) and MRC (\cite{mcc96}) surveys, and the USS samples from the 4C (\cite{cha96a}) and the one presented in this paper.
The results (Fig. \ref{zalpha}) show a trend for steeper spectral index sources to have higher redshifts in flux limited, spectrally unbiased samples, confirming the empirical relation out to the highest redshifts. The efficiency of the USS criterion is clearly illustrated by the fact that the 4C USS sample (\cite{cha96a}) contains 50\% $z>2$ sources, and by the early spectroscopic results on the USS samples presented in this paper, which indicate that $\sim$2/3 of our sources have $z>2$. It is even more impressive to note that 13 of the 14 radio galaxies at $z>3.5$ we know of have been found from samples with a steep spectral index selection\footnote{The only exception is VLA J123642+6213 (\cite{wad99}), which has been identified in the HDF, but it does have a steep spectral index ($\alpha_{1400}^{8500}=-0.94$).}! The limitation of this technique is that the steepest spectrum sources are rare, comprising typically only 0.5\% (at $\alpha < -1.30$) of a complete low frequency sample; therefore, large and deep all sky surveys are needed to obtain a significant sample of USS sources.

With the advent of several new deep all-sky surveys (\S 2), it is now possible for the first time to construct a well defined all-sky USS sample with optimized selection criteria to find large numbers of $z>3$ radio galaxies. In this paper, we describe the construction of such a sample, and present high resolution radio observations needed to determine accurate positions and morphologies. This information is essential for the optical and near-IR identifications, and subsequent optical spectroscopy of a significant sub-set of our sample, which will be described in a future papers. 
The organization of the paper is as follows: we describe the radio surveys we used in \S 2 and define our samples in \S 3. We present and discuss our radio observations in \S 4, and present our conclusions in \S 5.  
\begin{table*}
\centerline{\bf Table 1: Radio Surveys}

\footnotesize
\begin{tabular}{lccc}
\hline
\\
 & WENSS & TEXAS & MRC \\
\\
\hline
\\
Frequency (MHz) & 325 & 365 & 408 \\
Sky region (J2000) & $\delta >$ +29\arcdeg & $-$35\fdg7 $< \delta <$ +71\fdg5 & $-$85\arcdeg $< \delta <$ +18\fdg5 \\
\# of sources & 229,576 & 67,551 & 12,141 \\
Resolution & $ 54\arcsec \times$ 54\arcsec cosec$\delta$ & 10\arcsec$^a$ & $2\farcm62 \times 2\farcm86 \sec (\delta - 35\fdg5)$ \\
Position uncertainty & 1\farcs5 & 0\farcs5---1\arcsec & 8\arcsec \\
(strong sources) & & & \\
RMS noise & $\sim$4 mJy & 20 mJy & 70 mJy \\
Flux density limit & 18 mJy & 150 mJy & 670 mJy \\
Reference & \cite{ren97} & \cite{dou96} & \cite{lar81} \\
\\
\hline
\hline
\\
 & NVSS & FIRST & PMN \\
\\
\hline
\\
Frequency (MHz) & 1400 & 1400 & 4850 \\
Sky region (J2000) & $\delta > -$40\arcdeg$^b$ & $7^h20^m < \alpha < 17^h20^m$; +22\fdg2 $< \delta <$ +57\fdg5 & $-$87\fdg5 $< \delta <$ +10\arcdeg \\
 & & $21^h20^m < \alpha < 3^h20^m$; $-$2\fdg5 $< \delta <$ +1\fdg6 \\
\# of sources & 1,689,515 & 437,429 & 50,814 \\
Resolution & 45\arcsec $\times$ 45\arcsec & 5\arcsec $\times$ 5\arcsec & 4\minpoint2 \\
Position uncertainty & 1\arcsec & 0\farcs1 & $\sim$45\arcsec \\
(strong sources) & & & \\
RMS noise & 0.5 mJy & 0.15 mJy & $\sim$8 mJy \\
Flux density limit & 2.5 mJy & 1 mJy & 20 mJy \\
Reference & \cite{con98} & \cite{bec95} & \cite{gri93} \\
\hline
\end{tabular}

$^a$ The Texas interferometer has a complicated beam. However, sources with separations between 10\arcsec\ and 2\arcmin\ can be successfully modeled as doubles, and will have a single entry in the catalog. See \cite{dou96} for details.

$^b$  Some small gaps are not covered. They are listed on the NVSS homepage (1998 January 19 version).
\end{table*}

\section{Description of the Radio Surveys}

\begin{figure}[t]
\psfig{file=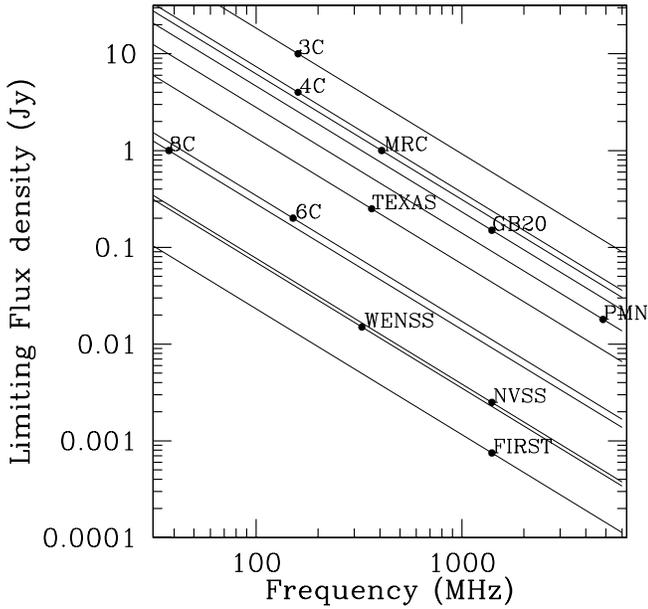,width=8.6cm}
\caption[surveylimits.ps]{Limiting flux density plotted for all major radio surveys. Lines are of constant spectral indices of $-1.3$. Note that WENSS, NVSS and FIRST have flux density limits $\sim 100$ times deeper than previous surveys at comparable wavelengths. \label{surveylimits}}
\end{figure}

During the past years, several all-sky radio-surveys have become available (Table 1), which are 1--2 orders of magnitude more sensitive than previous surveys at similar frequencies (Fig.\ \ref{surveylimits}). The combination of these new surveys allows us to define for the first time a large sample of USS sources that covers the whole sky\footnote{To facilitate optical follow-up, we will exclude the Galactic plane at $|b|<15$\degr} in both hemispheres. We list the main survey parameters in Table 1. In this section, we will briefly discuss the usefulness of these new radio surveys for the construction of USS samples. 

\subsection{WENSS}
The Westerbork Northern Sky Survey (WENSS; \cite{ren97}) at 325~MHz is the deepest low-frequency survey with a large sky coverage (3.14 sr). We used the WENSS to define the largest, and most complete USS sample to date, covering the entire sky North of declination 29\arcdeg. We used version 1.0 of the main and polar WENSS catalogs. A small area is covered by both these catalogs; we selected only the sources from the main catalog in this overlapping area.

\subsection{Texas}
\begin{figure}
\psfig{file=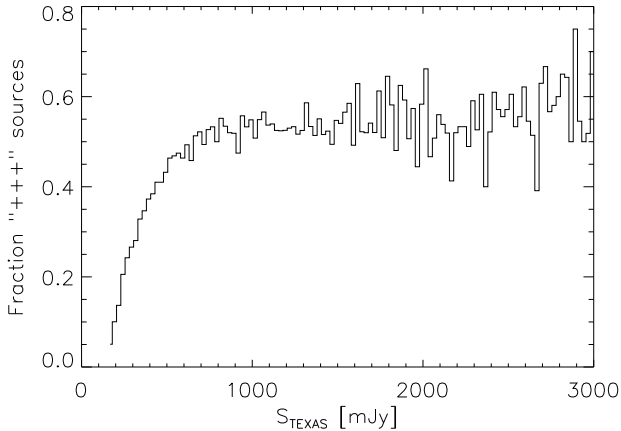,width=9cm}
\caption[texasppp.ps]{Fraction of sources with '+++'\ flag in the Texas catalog (see text) as a function of Texas $S_{365}$ flux density. Note that the selection of '+++' sources excludes primarily sources with $S_{365} \lesssim 700$~mJy. \label{texasppp}}
\end{figure}

The Texas survey, made with the Texas interferometer from 1974 to 1983 (\cite{dou96}), covers 9.63 steradians at a frequency of 365~MHz with a limiting flux density about ten times higher than that of the WENSS. The Texas interferometer's 3.5~km maximum baseline provides $<$1\arcsec\ positional accuracy, but its poor uv-coverage leads to irregular beamshapes and lobe-shifts, hampering accurate modeling of extended sources. A detailed discussion of these complications can be found in Douglas \etal\ (1996).
To minimize these problems, we have selected only the 40.9\% sources that are well modeled (listed with a '+++' flag in the catalog). This selection excludes primarily $S_{365} \lesssim 700$~mJy sources (Fig. \ref{texasppp}), but even at $S_{365} \gtrsim 700$~mJy, one out of three sources is excluded by this criterion.
\cite{dou96} have calculated the completeness above flux density S of the Texas catalog (defined as the fraction of sources with true flux density greater than S which appear in the catalog) by comparing the Texas with the MRC (\cite{lar81}) and a variety of other low-frequency catalogs. They found that the completeness varies with declination (because the survey was done in declination strips over a large time span), and an expected increase in completeness at higher flux densities. In Table 2, we reproduce their completeness table, extended with the values after the '+++' selection.

\begin{table}
\centerline{\bf Table 2: Completeness of the Texas survey}
\begin{tabular}{lcc}
\hline
Limiting flux density & all sources & '+++' sources \\
\hline
250 mJy & 0.8  & 0.2  \\
350 mJy & 0.88 & 0.28 \\
500 mJy & 0.92 & 0.40 \\
750 mJy & 0.96 & 0.51 \\
1 Jy & 0.96 & 0.50 \\
\hline
\end{tabular}
\end{table}

\begin{figure}
\psfig{file=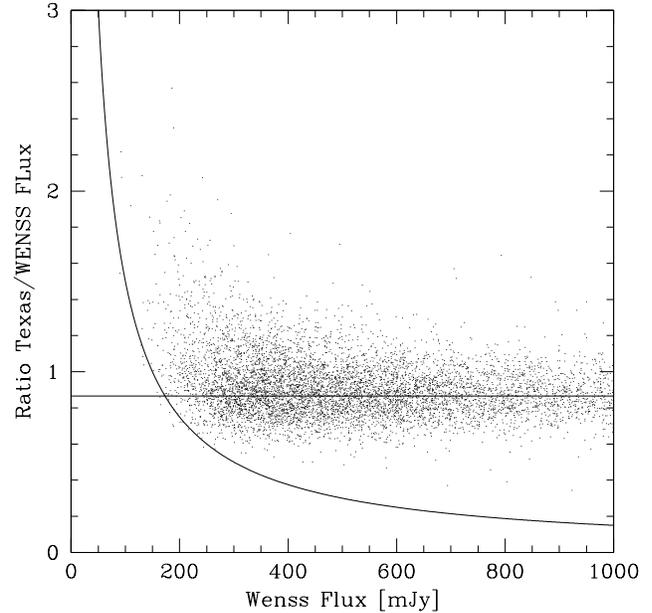,width=8.6cm}
\caption[wtratio.ps]{Ratio of integrated Texas over WENSS flux density against integrated WENSS flux density. Only Texas '+++' sources are plotted. The horizontal line is the expected ratio 0.865 due to the 40 MHz difference in central frequency, and assuming a spectral index of $\alpha_{365}^{1400}=-0.88$. The curved line indicates a 150 mJy flux density limit in the Texas catalog. Note the increasing amount of overestimated Texas flux densities with decreasing flux density $S_{WENSS} < 400$~mJy. \label{wtratio}}
\end{figure}

To examine the reliability of the listed flux densities, and to check to what extent the '+++' selection has removed the spurious sources from the catalog, we have correlated the Texas '+++' sources with WENSS, NVSS and FIRST.
In Figure \ref{wtratio}, we compare the Texas flux densities with those of the WENSS. At $S_{325} \gtrsim 500$~mJy, the ratio of the flux densities is closely distributed around 0.9. This ratio is what we expect due to the 40 MHz central frequency difference between the two surveys and assuming a spectral index $\alpha_{365}^{1400} = -0.879$ (the median of the Texas-NVSS spectral indices).
At $S_{325} \lesssim 500$~mJy, the number of sources in the Texas catalog which are brighter than in WENSS catalog increases with decreasing flux density. This can be explained by the 'up-scattering' of sources near the flux limit of the Texas catalog (\ie only sources intrinsically brighter than $S_{365} = 150$~mJy will be detected, but no $S_{365} < 150$~mJy sources with a large positive flux density measurement error). The result of this on a USS sample based on the Texas survey and correlated with a higher frequency survey (such as the NVSS), will be that with lower $S_{365}$, we will find more sources whose spectral indices appear steeper than they really are.

We also examined the dependence of the ratio Texas/WENSS flux density on angular size, determined from the FIRST survey (see \S 2.1.4). We found no significant residual variation of the flux density ratio at sizes between 5\arcsec\ and 2\arcmin.

In Figure \ref{usssr}b, we plot the density of NVSS sources around a Texas source (see also \S 2.2.5). The width of the over-density peak ($\sim 10$\arcsec) is due to the positional inaccuracies in the Texas and NVSS catalogues. However, the very broad tail of sources between 20\arcsec\ and 110\arcsec\ and the secondary peak coinciding with the fringe separation at 73\arcsec\ indicates that the '+++' selection did not remove all spurious sources from the catalog.

In summary, after the selection of '+++' sources, the Texas catalog still contains $<5$\% spurious sources (\cite{dou96}), probably due to residual lobe-shifted sources. Our comparison of the Texas flux densities with those of the WENSS survey shows that the differences are consistent with the errors quoted in the catalog. The selection of the Texas catalog with only '+++' sources is thus $>$95\% reliable, but only $\sim$40\% complete.

\subsection{NVSS}
The NRAO VLA Sky Survey (NVSS; \cite{con98}) covers the 10.3 steradians north of $-$40\arcdeg\ at 1.4~GHz, and reaches a 50 times lower limiting flux density than previous large area 1.4~GHz surveys. At the flux density levels we are using ($S_{1400}>10$~mJy), the catalog is virtually complete.
Because the NVSS resolution is comparable to that of the WENSS and Texas surveys, and its sky coverage is large, we use the NVSS to determine the spectral indices in our USS samples based on the WENSS and Texas surveys.
The final NVSS catalog was not yet completed at the time of our USS sample construction. For our final sample, presented in this paper, we use the 1998 January 19 version. This version still lacks data in a small number of regions of the sky (listed on the NVSS homepage). As a result, the sky coverage of the area listed in Table 3 is only 99.77\%.

\subsection{FIRST}
The Faint Images of Radio Sky at Twenty centimeters (FIRST, \cite{bec95}) survey is currently being made with the VLA in the B-array at 1.4~GHz, and has a limiting flux density three times deeper than the NVSS. We used the 1998 February 4 version of the catalog, covering 1.45~steradians. As noted by Becker \etal\ (1995), the photometry for extended sources in FIRST might be less reliable than that of the NVSS, due to the $9 \times$ higher resolution, which could underestimate large-scale diffuse radio emission. As the FIRST area is completely covered by the NVSS, we will consistently use NVSS flux densities for our spectral index calculation. The main advantages of FIRST over NVSS for our purposes are the much better positional accuracy ($<$~0\farcs5) and the higher (5\arcsec) resolution. This combination allows the identification of even the very faint ($R > 20$) optical counterparts of radio sources. Additionally, the fainter detection limit of the FIRST allows an extra check on the flux densities of compact sources.

\subsection{MRC}
The Molonglo Reference Catalog of radio sources (\cite{lar81}) at 408~MHz is presently the most sensitive low-frequency catalog with reasonable positional accuracy that covers the deep southern hemisphere, $\delta < -35$\arcdeg. We will use this catalog in combination with the PMN survey (see below) to define the first USS sample at $\delta < -40\arcdeg$.

\subsection{PMN}
The Parkes-MIT-NRAO (PMN) survey is a combination of 4 strips observed with the Parkes telescope at 4.85~GHz. The strips cover different parts of the sky, each with a slightly different limiting flux density. The regions are: southern ($-87\fdg5 < \delta < -37\arcdeg$, \cite{wri94}), zenith ($-37\arcdeg < \delta < -29\arcdeg$, \cite{wri96}), tropical ($-29\arcdeg < \delta < -9\fdg5$, \cite{gri94}), and equatorial ($-9\fdg5 < \delta < +10\arcdeg$, \cite{gri95}). For our southern hemisphere sample, we have used the southern and zenith catalogs to find USS sources at $\delta < -30\arcdeg$.

\begin{figure}
\psfig{file=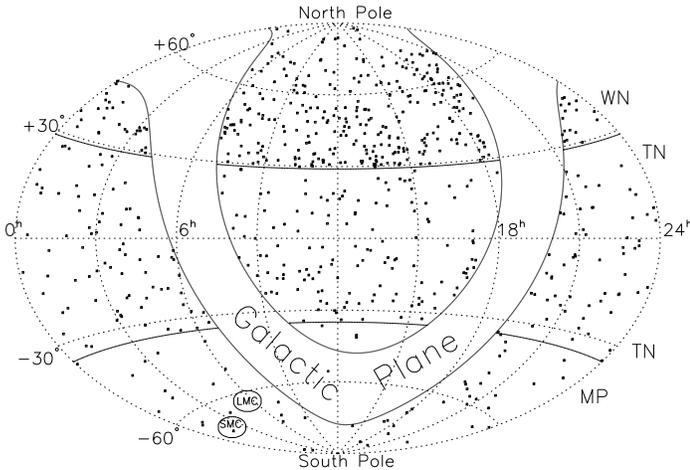,width=8.8cm}
\caption[usssky.ps]{Sky coverage of our 3 USS samples. Constant declination lines denote the boundaries between our the WN and TN and between the TN and MP samples, as indicated on the right. Note the difference in source density and the exclusion of the Galactic Plane. \label{usssky}}
\end{figure}

\section{USS Samples}

\begin{figure*}
\psfig{file=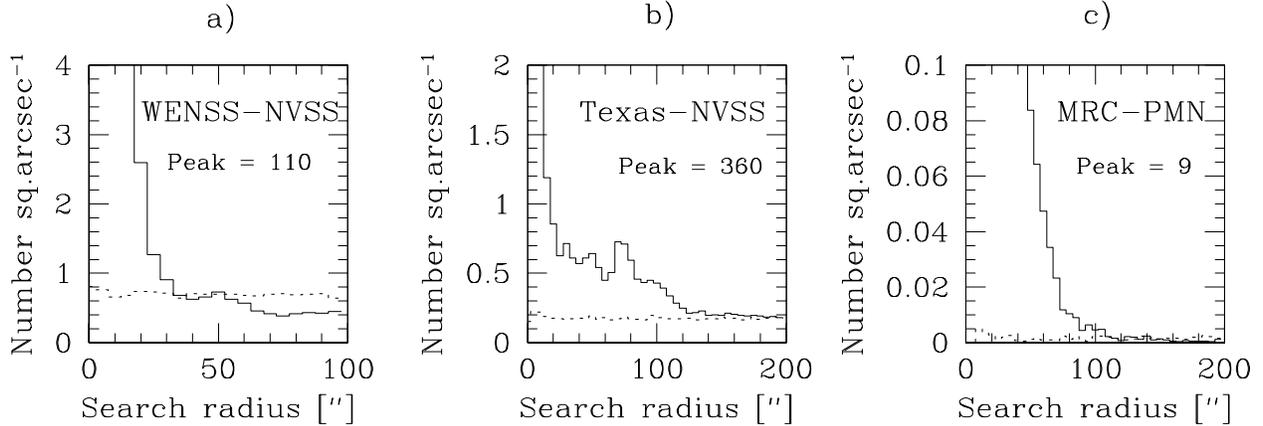,width=18cm}
\caption[usssr.ps]{The density of sources from the high frequency catalog used in the correlation (NVSS or PMN) around sources from the low frequency catalog (WENSS, Texas or MRC) as a function of search radius. The dotted line represents the distribution of confusion sources (see text). The apparent under-density of WENSS--NVSS sources at search radii $\simgt 60\arcsec$ is due to the grouping of multiple component sources in the $\sim$1\arcmin\ resolution WENSS. Note the plateau and secondary peak around the 73\arcsec\ Texas fringe separation in the Texas--NVSS correlation and the much larger uncertainty in the MRC--PMN correlation.\label{usssr}}
\end{figure*}

Figure \ref{surveylimits} shows that the surveys described above have very compatible flux density limits for defining samples of USS sources. At the same time, their sky coverage is larger and more uniform than previous surveys used for USS sample construction (\cite{wie92}, \cite{rot94}; \cite{cha96a}, \cite{blu98}; \cite{ren98}, \cite{pur99}, \cite{ped99}, \cite{and00}). We selected the deepest low and high frequency survey available at each part of the sky. For a small region $-35\arcdeg < \delta < -30\arcdeg$ which is covered by both Texas and MRC, we used both surveys. This resulted in a more complete samples since the lower sensitivity of the PMN survey in the zenith strip (see \S2.6) is partly compensated by the (albeit incomplete) Texas survey.
To avoid problems with high Galactic extinction during optical imaging and spectroscopy, all regions at Galactic latitude $|b| <$ 15\arcdeg\ were excluded\footnote{This also reduces the number of Galactic pulsars in our sample (see \S4.7.2).}, as well as the area within 7\arcdeg of the LMC and SMC. This resulted in three USS samples that cover a total of 9.4 steradians (Fig. \ref{usssky}).

We designate the USS samples by a two-letter name, using the first letter of their low- and high-frequency contributing surveys. Sources from these samples are named with this 2-letter prefix followed by their IAU J2000-names using the positions from the NVSS catalog (WN and TN samples) or the MRC catalog (MP sample). We did not rename the sources after a more accurate position from our radio observations or from the FIRST survey. The sample definitions are summarized in Table 3.

\subsection{Survey combination issues}
We first discuss the problems that arise when combining radio surveys with different resolutions and positional uncertainties.
\subsubsection{Correlation search radius}
Due to the positional uncertainties and resolution differences between radio surveys, in general the same source will be listed with slightly different positions in the catalogs.

To empirically determine the search radius within which to accept sources in 2 catalogs to be the same, we compared the density of objects around the position listed in the low-frequency survey (which has lower resolution) with the expected number of random correlations in each sample ($\equiv$ confusion sources). To determine this number as a function of distance from the position in the most accurate catalog, we created a random position catalog by shifting one of the input catalogs by 1\arcdeg\ in declination, and made a correlation with this shifted catalog. The density of sources as a function of distance from the un-shifted catalog then represents the expected number of confusing sources as a function of radial distance. In Figure \ref{usssr}, we plot for each of our three samples the observed density around these sources with this confusing distribution over-plotted. The correlation search radius should thus be chosen at a distance small enough for the density of confusion sources to be negligible.

We decided to adopt the radius where the density of real sources is at least ten times higher that the density of confusion sources as the search radius for our sample construction, except for the WN sample (would be 15\arcsec) where we chose the same radius as for the TN sample (10\arcsec). The later was done for consistency between both samples. Because of the five times lower resolution and source densities in the MRC and PMN surveys, the search radius of the MP sample is eight times larger. Summarized, the search radii we used are 10\arcsec\ for WN and TN, and 80\arcsec\ for MP.
\begin{table*}
\centerline{\bf Table 3: USS samples}
\scriptsize
\begin{tabular}{lrccccccc}
\hline
Sample & Sky Area$\qquad\quad$ & Density & Spectral Index & Search Radius & Flux Limit & C$^a$ & R$^a$ & \# of Sources \\
 &  & sr$^{-1}$ & & & mJy & &\\
\hline
WN & 29\arcdeg $< \delta <$ 75\arcdeg, $|b| >$ 15\arcdeg$^b$ & 151 & $\alpha_{325}^{1400} \le -1.30$ & 10\arcsec & $S_{1400} >$ 10 & 96\% & 90\% & 343 \\
TN & $-$35\arcdeg $< \delta <$ 29\arcdeg, $|b| >$ 15\arcdeg$^b$ & 48$^c$ & $\alpha_{365}^{1400} \le -1.30$ & 10\arcsec & $S_{1400} >$ 10 & 97\%$^c$ & 93\% & 268 \\
MP & $\delta < -$30\arcdeg, $|b| >$ 15\arcdeg & 26 & $\alpha_{408}^{4800} \le -1.20$ & 80\arcsec & S$_{408} > 700$; S$_{4850} > 35$ & 100\% & 100\% & 58 \\
\hline
\end{tabular}

$^a$ C=completeness and R=reliability accounting only for scattering across the spectral index limit (see \S3.3.1).

$^b$ coverage is only 99.7\% because some small patches of sky we not covered at the time of writing. They are listed on the NVSS homepage \\(1998 January 19 version).

$^c$ Because we selected only problem free sources from the Texas survey, the effective completness of the TN sample is $\sim 30\%$.
\end{table*}

\subsubsection{Angular size}
In order to minimize errors in the spectral indices due to different resolutions and missing flux on large angular scales in the composing surveys, we have only considered sources which are not resolved into different components in the composing surveys. Effectively, this imposes an angular size cutoff of $\sim$1\arcmin\ to the WN, $\sim$2\arcmin\ to the TN sample and $\sim$4\arcmin\ to the MP sample.
We deliberately did not choose a smaller angular size cutoff (as \eg Blundell \etal\ (1998) did for the 6C$^*$ sample), because (1) higher resolution angular size information is only available in the area covered by the FIRST survey, and (2) even a 15\arcsec\ cutoff would only reduce the number of sources by 30\%, while it would definitely exclude several HzRGs from the sample. For example, in the 4C USS sample (\cite{cha96b}), three out of eight $z>2$ radio galaxies have angular sizes $>$15\arcsec.

We think that our $\sim$1\arcmin\ angular size cutoff will exclude almost no HzRGs, because the largest angular size for $z > 2$ radio galaxies in the literature is 53\arcsec\ (4C~23.56 at $z = 2.479$; \cite{cha96a}, \cite{car97}), while all 45 $z > 2.5$ radio galaxies with good radio maps are $<$~35\arcsec\ (\cite{car97}). Although the sample of known $z>2$ radio galaxies is affected by angular size selection effects, very few HzRGs larger than 1\arcmin\ would be expected.

The main incompleteness of our USS sample stems from the spectral index cutoff and the flux limit (\S3.2). However, our flux limit ($S_{1400}$=10~mJy) is low enough to break most of the redshift-radio power degeneracy at $z>2$. To achieve this with flux limited samples, multiple samples are needed (\eg \cite{blu99}).

\subsection{Sample definition}

\subsubsection{WENSS-NVSS (WN) sample}
A correlation of the WENSS and NVSS catalogs with a search radius of 10\arcsec\ centered on the WENSS position (see \S 3.1.1) provides spectral indices for $\sim 143,000$ sources. Even with a very steep $\alpha_{325}^{1400} \le -1.30$ spectral index criterion, we would still have 768 sources in our sample.
To facilitate follow-up radio observations, and to increase the accuracy of the derived spectral indices (see \S 3.3.1), we have selected only NVSS sources with $S_{1400} > 10$~mJy.  Because the space density of the highest redshift galaxies is low, it is important not to limit the sample area (see \eg \cite{raw98}) to further reduce the number of sources in our sample.
Because the NVSS has a slightly higher resolution than the WENSS (45\arcsec\ compared to $54\arcsec \times 54\arcsec $cosec$~\delta$), some WENSS sources have more than one associated NVSS source. We have rejected the 11 WN sources that have a second NVSS source within one WENSS beam. Instead of the nominal WENSS beam ($54\arcsec \times 54\arcsec$cosec$\delta$), we have used a circular 72\arcsec\ WENSS beam, corresponding to the major axis of the beam at $\delta = 48\arcdeg$, the position that divides the WN sample into equal numbers to the North and South. 
The final WN sample contains 343 sources.

\subsubsection{Texas-NVSS (TN) sample}
Because the Texas and NVSS both have a large sky-coverage, the area covered by the TN sample includes 90\% of the WN area. In the region $\delta >29\arcdeg$, we have based our sample on the WENSS, since it does not suffer from lobe-shift problems and reaches ten times lower flux densities than the Texas survey (\S 2.2).
In the remaining 5.28 steradians South of declination +29\arcdeg, we have spectral indices for $\sim 25,200$ sources. Again, we used a 10\arcsec\ search radius (see \S 3.1.1), and for the same reason as in the WN sample we selected only NVSS sources with $S_{1400} > 10$~mJy. Combined with the $\alpha_{365}^{1400} \le -1.30$ criterion, the number of USS TN sources is 285.
As for the WN sample, we further excluded sources with more than one $S_{1400}>$10~mJy NVSS source within 60\arcsec\ around the TEXAS position, leaving 268 sources in the final TN sample. We remind (see \S 2.2) that the selection of the TEXAS survey we used is only $\sim$40\% complete with a strong dependence on flux density. Using the values from table 2, we estimate that the completeness of our TN sample is $\sim$30\%.

\subsubsection{MRC-PMN (MP) sample}
In the overlapping area, we preferred the TN over the MP sample for the superior positional accuracies and resolutions of both Texas and NVSS compared to MRC or PMN. Because the MRC survey has a low source density, we would have only 13 MP sources with $\alpha_{408}^{4850} \le -1.30$. We therefore relaxed this selection criterion to $\alpha_{408}^{4850} \le -1.20$, yielding a total sample of 58 sources in the deep South ($\delta < -30$\arcdeg).
\begin{figure*}
\centerline{\psfig{file=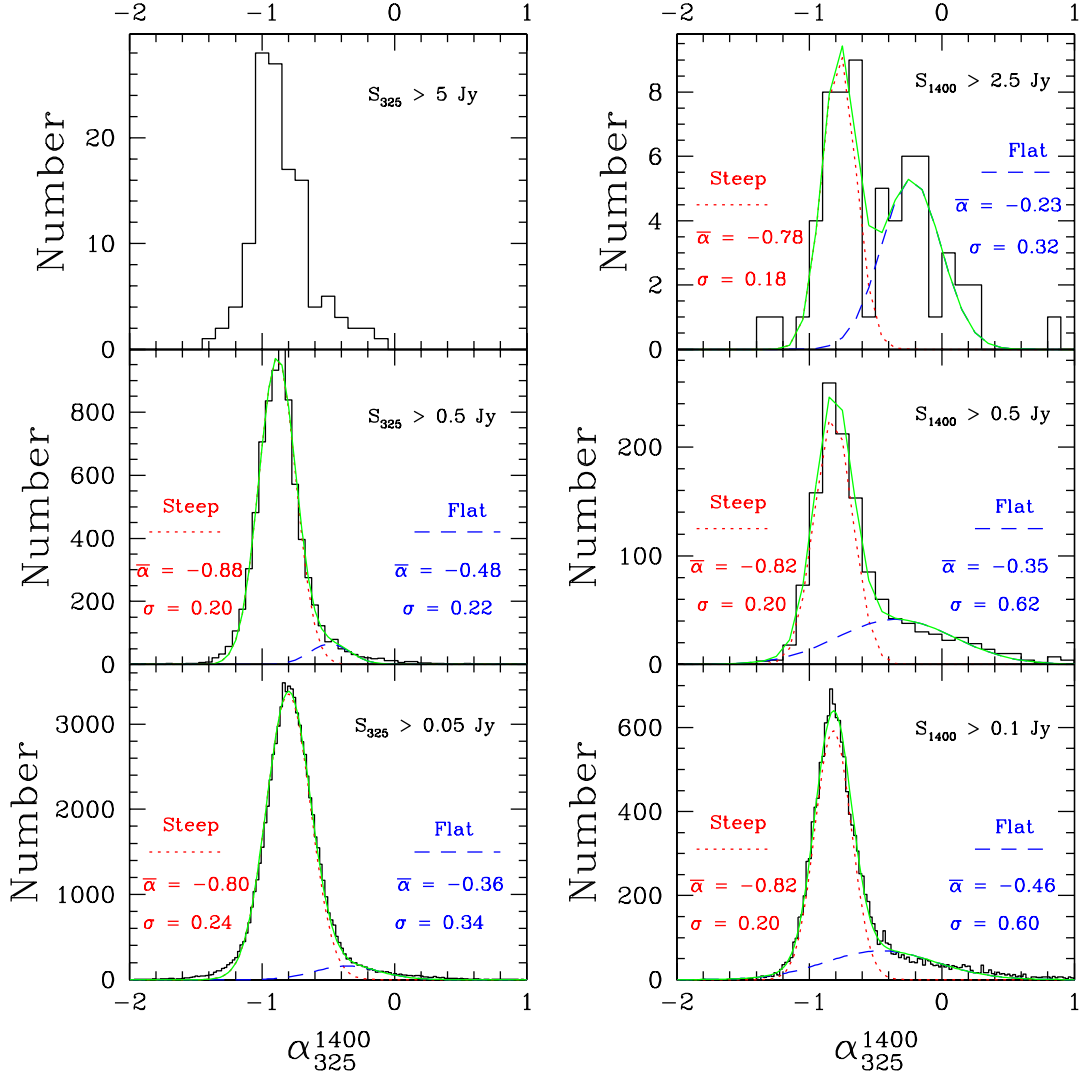,width=15cm}}
\caption[wn3his.ps]{Spectral index distributions from the WENSS-NVSS correlation. The left and right panels show the variation with 325~MHz and 1.4~GHz flux density. A low frequency selected sample is more appropriate to study the steep-spectrum population. The parameters of a two-component Gaussian fit (dotted line = steep, dashed line = flat) are shown in each panel. The solid line is the sum of both Gaussians. \label{wn3spixhis}}
\end{figure*}

\subsection{Discussion}

\subsubsection{Spectral index errors}
We have listed the errors in the spectral indices due to flux density errors in the catalogs in Tables A.1 to A.3. The WN and TN samples have the most accurate spectral indices: the median spectral index errors are $\Delta\bar{\alpha}_{325}^{1400} = 0.04$ for WN sources and $\Delta\bar{\alpha}_{365}^{1400} = 0.04$ ($S_{365} >$ 1~Jy) to 0.07 ($S_{365} >$ 150~mJy) for TN sources. For the MP sample, $\Delta\bar{\alpha}_{408}^{4850} \approx 0.1$, with little dependence on flux density (S$_{408} > 750$mJy).

\begin{figure}
\psfig{file=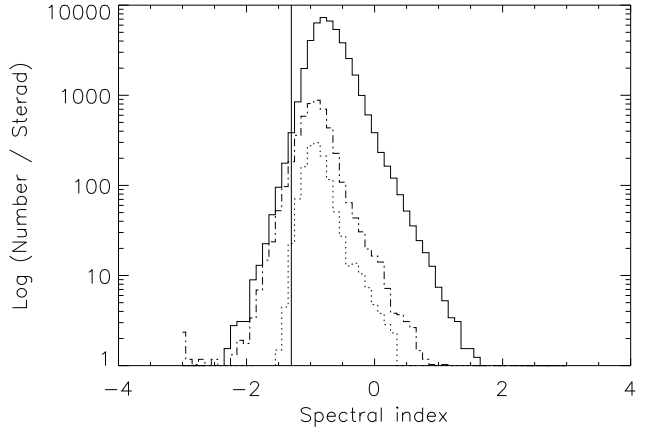,width=9cm}
\caption[spixhis.ps]{ Logarithmic spectral index distribution for WENSS-NVSS (full line), Texas-NVSS (dot-dash line) and MRC-PMN (dotted line). The vertical line indicates the -1.3 cutoff used in our spectral index selection. Note the difference in number density and the sharper fall-off on the flat-end part of the TN and MP compared to the WN. \label{spixhis}}
\end{figure}

Because our sample selects the sources in the steep tail of the spectral index distribution (Fig. \ref{wn3spixhis} and \ref{spixhis}), there will be more sources with an intrinsic spectral index flatter than our cutoff spectral index that get scattered into our sample by measurement errors than there will be sources with intrinsic spectral index steeper than the cutoff that get scattered out of our sample. 

Following the method of \cite{ren98}, we fitted the steep tail between $-1.60 < \alpha < -1.0$ with a Gaussian function. For each of our three samples, we generated a mock sample drawn from this distribution, and added measurement errors by convolving this true spectral index distribution with a Gaussian distribution with as standard deviation the mean error of the spectral indices. The WN mock sample predicts that 13 $\alpha_{325}^{1400} < -1.30$ sources get scattered out of the sample while 36 $\alpha_{325}^{1400} > -1.30$ sources get scattered into the USS sample.
Thus, the WN sample is 96\% complete and 90\% reliable. For the TN sample, we expect to loose 7 $\alpha_{365}^{1400} < -1.30$ sources\footnote{only due to the spectral index cutoff, the sample has more important incompleteness factors; see \S 2.2}, and have 18 contaminating $\alpha_{365}^{1400} > -1.30$ sources. The completeness is thus 97\% and the reliability 93\%. For the MP sample, this spectral index scattering is negligible, because there are too few sources in the steep spectral index tail.

Our reliability and completeness are significantly better than the values of $\sim$75\% and $\sim$50\% of \cite{ren98} because (1) our spectral indices are more accurate because they were determined from a wider frequency interval than the 325--610~MHz used by \cite{ren98}, and (2) our sample has a steeper cutoff spectral index, where the spectral index distribution function contains fewer sources and has a shallower slope, leading to fewer sources that can scatter in or out of the sample.

\subsubsection{Spectral index distributions}
Using the 143,000 spectral indices from the WENSS-NVSS correlation, we examined the flux density dependence of the steep and flat spectrum sources. Selecting sources with $S_{325}>50$~mJy or $S_{1400}>100$~mJy assures that we will detect all sources with $\alpha_{325}^{1400}>{\frac{\ln(S_{NVSS}^{lim}/50)}{\ln(325/1400)}}=-1.82$ or $\alpha_{325}^{1400}<{\frac{\ln(S_{WENSS}^{lim}/100)}{\ln(325/1400)}}=0.82$ respectively, where $S_{NVSS}^{lim}=3.5$~mJy  and $S_{WENSS}^{lim}=30$~mJy are the lowest flux densities where the NVSS and WENSS are complete (\cite{con98}, \cite{ren97}). The results shown in Figure \ref{wn3spixhis} therefore reflect only the effect of a different selection frequency. Two populations are present in both the $S_{325}$ and $S_{1400}$ selected distributions. The peaks of the steep and flat populations at $\bar{\alpha}_{325}^{1400} \approx -0.8$ and $\bar{\alpha}_{325}^{1400} \approx -0.4$ do not show significant shifts over three orders of magnitude in flux density. This is consistent with the results that have been found at 4.8~GHz (\cite{wit79}, \cite{mac81}, \cite{owe83}), with the exception that their $\bar{\alpha}_{1400}^{4800} \approx 0.0$ for the flat spectrum component is flatter than the $\bar{\alpha}_{325}^{1400} \approx -0.4$ we found. However, we find that the relative contribution of the flat spectrum component increases from 25\% at $S_{1400}>0.1$~Jy to 50\% at $S_{1400}>2.5$~Jy

Because the steep- and flat-spectrum populations are best separated in the $S_{1400}>$2.5~Jy bin, we have searched the literature for identifications of all 58 $S_{1400} > 2.5$~Jy sources to determine the nature of both populations. All but one (3C 399, \cite{mar98}) of the objects outside of the Galactic plane ($|b| > 15\arcdeg$) were optically identified. Of the 30 steep spectrum ($\alpha_{325}^{1400} < -0.6$) sources, two thirds were galaxies, while the rest were quasars. Half of the flat spectrum ($\alpha_{325}^{1400} > -0.6$) sources were quasars, 20\% blazars, and 30\% galaxies. Figure \ref{wn3spixhis} therefore confirms that the steep and flat spectral index populations are dominated by radio galaxies and quasars respectively. We also find that while the relative strength between the steep and flat spectrum populations changes due to the selection frequency, the median spectral index and width of the population does not change significantly over three orders in magnitude of flux density. Even fainter studies would eventually start to get contamination from the faint blue galaxy population (see \eg\ \cite{win85}).

\begin{table*}
\centerline{\bf Table 4: Radio Observations}
\begin{center}
\begin{tabular}{lrrrcr}
\hline
UT Date & Telescope & Config. & Frequency & Resolution & \# of sources \\
\hline
1996 October 28 & VLA & A & 4.86 GHz & 0\farcs3 & 90 WN, 25 TN \\
1997 January 25 & VLA & BnA& 4.86 GHz & 0\farcs6 & 29 TN \\
1997 March 10 & VLA & BnA & 4.885 GHz & 0\farcs6 & 8 TN \\
1997 December 15 & ATCA & 6C & 1.420 GHz & 6\arcsec $\times$ 6\arcsec\ cosec$\delta$ & 41 MP, 32 TN \\
1998 August 12+17 & VLA & B & 4.86 GHz & 1\farcs0 & 151 WN \\
\hline
\end{tabular}
\end{center}
\end{table*}

\subsubsection{Consistency of the three USS samples}
We compare the spectral index distributions of our three USS samples in logarithmic histograms (Fig. \ref{spixhis}). The distributions are different in two ways. First, the WENSS-NVSS correlation contains nine times more sources than the Texas-NVSS, and 14 times more than the MRC-PMN correlation. Second, the shapes of the distributions are different: while the steep side of the TN sample coincides with that of the WN, its flat end part falls off much faster. The effect is so strong that it even shifts the TN peak steep-wards by $\sim0.15$. For the MP sample, the same effect is less pronounced, though still present.

Both effects are due to the different flux density limits of the catalogs. The deeper WENSS catalog obviously contains more sources than the TEXAS or MRC catalogs, shifting the distributions vertically in Figure \ref{surveylimits}. The relative 'shortage' of flat spectrum sources in the Texas-NVSS and MRC-PMN correlations can be explained as follows. A source at the flux density limit in both WENSS and NVSS would have a spectral index of $\alpha_{325}^{1400} = -1.3$, while for Texas and NVSS this would be $\alpha_{365}^{1400} = -1.7$ (see Fig. \ref{surveylimits}). Faint NVSS sources with spectral indices flatter than these limits will thus more often get missed in the TEXAS catalog than in the WENSS catalog. This effect is even strengthened by the lower completeness at low flux densities of the Texas catalog. However, very few USS sources will be missed in either the WENSS-NVSS or Texas-NVSS correlations\footnote{The TN sample will have more spurious sources at low flux density levels; see \S2.2}. The parallel slope also indicates that the USS sources from both the WENSS-NVSS and Texas-NVSS correlations were drawn from the same population of radio sources. We therefore expect a similar efficiency in finding HzRGs from both samples.

The MP sample has been defined using a spectral index with a much wider frequency difference. However, the observed ATCA 1.420 GHz flux densities can be used to construct $\alpha_{408}^{1420}$. An 'a posteriori' selection using $\alpha_{408}^{1420} \le -1.30$ from out ATCA observations (see \S 4.2) would keep $\sim$60\% of the MP sources in a WN/TN USS sample.

\section{Radio Observations}
Of all the major radio surveys described in \S 2, only FIRST has sufficient positional accuracy and resolution for the optical identification of $R > 20$ objects. We present FIRST maps of 139 WN and 8 TN sources in appendices B.2 and B.4.

Outside the area covered by FIRST, we have observed all the remaining WN sources, 30\% of the TN sample, and 71\% of the MP sources at 0\farcs3 to 5\arcsec\ resolution using the Very Large Array (VLA; \cite{nap83}) and Australia Telescope Compact Array (ATCA; \cite{fra92}) telescopes.
A log of the radio observations is given in Table 4. We observed targets for our VLA runs on the basis of declination (A-array for $\delta >$ 0\arcdeg\ and BnA-array for $\delta <$ 0\arcdeg) and sky coverage of the WN and TN samples, which were still incomplete at the time of the 1996 observations. We observed all WN, and most TN sources  with the VLA, and all MP sources with the ATCA. We observed TN sources between $-31$\arcdeg $< \delta < -10$\arcdeg\ with either VLA or ATCA, depending on the progress of the NVSS at the time of the observations.

\subsection{VLA observations and data reduction}

We observed all sources in the standard 4.86 GHz C-band with a 50 MHz bandwidth, resulting in a resolution of $\sim$0\farcs3 in the A-array and $\sim$1\arcsec\ in the BnA-array. We spent 5 minutes on each source, implying a theoretical rms level of 75 $\mu$Jy, or a ratio of total integrated signal over map noise of 110 for the weakest sources, assuming no spectral curvature beyond 1.4 GHz.
We performed calibration and data editing in \aips, the Astronomical Image Processing System from NRAO.
We used 3C286 as the primary flux calibrator in all runs. Comparison of the flux density of 3C48 with the predicted values indicated the absolute flux density scale was accurate up to 2\%. We observed nearby (within 15\arcdeg) secondary flux calibrators every 15 to 20 minutes to calibrate the phases. After flagging of bad data, we spilt the uv-data up into separate data sets for imaging and self-calibration in DIFMAP, the Caltech difference mapping program (\cite{she97}).
We used field sizes of 164\arcsec\ (A$-$array) or 256\arcsec\ (BnA$-$array) with pixel scales of 0\farcs08 / pixel (A$-$array) or 0\farcs25 / pixel (BnA$-$array). Even the smallest field of view is still four times larger than the resolution of the NVSS, so all components of an unresolved NVSS source will be covered.

We cleaned each source brighter than the 5$\sigma$ level, followed by a phase-only self-calibration. We repeated the latter for all sources in the field of a source. Next, we made a new model from the (self-calibrated) uv-data, and subsequently cleaned to the level reached before. The last stage in the mapping routine was a deep clean with a 1\% gain factor over the entire field. Most of the resulting maps have noise levels in the range 75 to 100 $\mu$Jy, as expected.

\subsection{ATCA observations and data reduction}
We used the ATCA in the 6C configuration, which has a largest baseline of 6km. We observed at a central frequency of 1.384 GHz, which was selected to avoid local interference. We used 21 of the 30 frequency channels that had high enough signal, which resulted in an effective central frequency of 1.420 GHz, with a 84 MHz bandwidth. In order to obtain a good uv-coverage, we observed each source eight to ten times for three minutes, spread in hour angle. The primary flux calibrator was the source 1934-648; we used secondary flux calibrators within 20\arcdeg\ of the sources to calibrate the phases.
We performed editing and calibration in \aips, following standard procedures. We made maps using the automated mapping/self-calibration procedure MAPIT in \aips. The resulting 1.420 GHz maps (Fig. \ref{mpatca}) have noise levels of $\sim$5 mJy.

\subsection{Results}
Of all 343 WN sources, 139 have FIRST maps (appendix B.2). All remaining 204 sources were observed, and 141 were detected. The remaining 30\% were too faint at 4.86~GHz to be detected in 5~min snapshots, because their high frequency spectral index steepens more than expected, or they were over-resolved. Because they are significantly brighter, all the observed 89 TN and 41 MP sources were detected.
We present contour maps of all the detections in Appendices B.1, B.3, B.5, and B.6 and list the source parameters in Tables A.1 to A.3.

We have subdivided our sources into 5 morphological classes, using a classification similar to that used by R\"ottgering \etal\ (1994). Note that this classification is inevitably a strong function of the resolution, which varies by a factor of 20 between the VLA A-array and the ATCA observations. 

We have determined the source parameters by fitting two-dimensional Gaussian profile to all the components of a source. The results are listed in Tables A.1 to A.3 which contain:

\begin{description}
\item[Col 1:]{Name of the source in IAU J2000 format. The 2-letter prefix indicates the sample: WN: WENSS--NVSS, TN: Texas--NVSS, MP: MRC--PMN.}
\item[Col 2:]{The integrated flux density from the low-frequency catalog.}
\item[Col 3:]{The integrated flux density at the intermediate frequency, determined from the NVSS for WN and TN, or from the 1.420~GHz ATCA observations for the MP sample.}
\item[Col 4:]{The integrated flux density at 4.86 GHz, determined from the VLA observations for WN and TN, and from the PMN survey for the MP sample.}
\item[Col 5:]{The lower frequency two-point spectral index. This is the spectral index used to define the WN and TN samples.}
\item[Col 6:]{The higher frequency two-point spectral index. This is the spectral index used to define the MP sample.}
\item[Col 7:]{Morphological classification code: single (S), double (D), triple (T) and multiple (M) component sources, and irregularly shaped diffuse (DF) sources.}
\item[Col 8:]{Largest angular size. For single component sources, this is the de-convolved major axis of the elliptical Gaussian, or, for unresolved sources (preceded with $<$), an upper limit is given by the resolution. For double, triple and multiple component sources, this is the largest separation between their components. For diffuse sources this is the maximum distance between the source boundaries defined by three times the map rms noise.}
\item[Col 9:]{De-convolved position angle of the radio structure, measured North through East}
\item[Col 10 -- 11:]{J2000 coordinates, determined from the map with position code listed in col. 12. The positions in the VLA and ATCA maps have been fitted with a single two-dimensional elliptical Gaussian. For double (D) sources, the geometric midpoint is given; for triples (T) and multiples (M), the core position is listed. For diffuse (DF) sources we list the center as determined by eye.}
\item[Col 12:]{Position code, indicating the origin of the morphological and positional data in column 7 to 11: A=ATCA, F=FIRST, M=MRC, N=NVSS, and V=VLA}.
\end{description}

\subsection{Notes on individual sources}
\noindent{\bf WN~J0043+4719:} The source 18\arcsec\ north of the NVSS position is not detected in the NVSS. This is therefore not a real USS source because the NVSS flux density was underestimated.

\noindent{\bf WN~J0048+4137:} Our VLA map probably doesn't go deep enough to detect all the flux of this source.

\noindent{\bf WN~J0727+3020:} The higher resolution FIRST map shows that both components of this object are indeed identified on the POSS, even though the NVSS position is too far off to satisfy our identification criterion.

\noindent{\bf WN~J0717+4611:} Optical and near-IR spectroscopy revealed this object as a red quasar at $z=1.462$ (\cite{deb98b}).

\noindent{\bf WN~J0725+4123:} The extended POSS identification suggest this source is located in a galaxy cluster.

\noindent{\bf WN~J0829+3834:} The NVSS position of this unresolved source is 7\arcsec\ ($3\sigma$) from the FIRST position, which itself is only at 2\arcsec\ from the WENSS position.

\noindent{\bf WN~J0850+4830:} The difference with the NVSS position indicates that our VLA observations are not deep enough to detect a probable north-eastern component.

\noindent{\bf WN~J0901+6547:} This 38\arcsec\ large source is over-resolved in our VLA observations, and probably even misses flux in the NVSS, and is therefore not a real USS source.

\noindent{\bf WN~J1012+3334:} The bend morphology and bright optical sources to the east indicate this object is probably located in a galaxy cluster.

\noindent{\bf WN~J1101+3520:} The faint FIRST component 20\arcsec\ north of the brighter Southern component is not listed in the FIRST catalog, but is within 1\arcsec\ of a faint optical object. This might be the core of a 70\arcsec\ triple source.

\noindent{\bf WN~J1152+3732:} The distorted radio morphology and bright, extended POSS identification suggest this source is located in a galaxy cluster.

\noindent{\bf WN~J1232+4621:} This optically identified and diffuse radio source suggest this source is located in a galaxy cluster.

\noindent{\bf WN~J1314+3515:} The diffuse radio source appears marginally detected on the POSS.

\noindent{\bf WN~J1329+3046A,B, WN~J1330+3037, WN~J1332+3009 \& WN~J1333+3037:} The noise in the FIRST image is almost ten times higher than average due to the proximity of the $S_{1400}$=15~Jy source 3C~286.

\noindent{\bf WN~J1330+5344:} The difference with the NVSS position indicates that our VLA observations are not deep enough to detect a probable south-eastern component.

\noindent{\bf WN~J1335+3222:} Although the source appears much like the hotspot of a larger source with the core 90\arcsec\ to the east, no other hotspot is detected in the FIRST within 5\arcmin.

\noindent{\bf WN~J1359+7446:} The extended POSS identification suggests this source is located in a galaxy cluster.

\noindent{\bf WN~J1440+3707:} The equally bright galaxy 30\arcsec\ south of the POSS identification suggests that this source is located in a galaxy cluster.

\noindent{\bf WN~J1509+5905:} The difference with the NVSS position indicates that our VLA observations are not deep enough to detect a probable western component.

\noindent{\bf WN~J1628+3932:} This is the well studied galaxy NGC~6166 in the galaxy cluster Abell 2199 (\eg \cite{zab93}.

\noindent{\bf WN~J1509+5905:} The difference with the NVSS position indicates that our VLA observations are not deep enough to detect a probable west-south-western component.

\noindent{\bf WN~J1821+3601:} The source 35\arcsec\ south-west of the NVSS position is not detected in the NVSS. This is therefore not a real USS source because the NVSS flux density was underestimated.

\noindent{\bf WN~J1832+5354:} The source 19\arcsec\ north-east of the NVSS position is not detected in the NVSS. This is therefore not a real USS source because the NVSS flux density was underestimated.

\noindent{\bf WN~J1852+5711:} The extended POSS identification suggests this source is located in a galaxy cluster.

\noindent{\bf WN~J2313+3842:} The extended POSS identification suggests this source is located in a galaxy cluster.

\noindent{\bf TN~J0233+2349:} This is probably the north-western hotspot of a 35\arcsec\ source, with the south-eastern component barely detected in our VLA map. 

\noindent{\bf TN~J0309-2425:} We have classified this source as a 13\arcsec\ double, but the western component might also be the core of a 45\arcsec\ source, with the other hotspot around $\alpha = 3^h9^m10^s, \delta=-24\arcdeg25\arcmin50\arcsec$.

\noindent{\bf TN~J0349-1207:} The core-dominated structure is reminiscent of the red quasar WN~J0717+4611.

\noindent{\bf TN~J0352-0355:} This is probably the south-western hotspot of a 30\arcsec\ source.

\noindent{\bf TN~J0837-1053:} Given the 10\arcsec\ difference between the positions of the NVSS and diffuse VLA source, this is probably the northern component of a larger source.

\noindent{\bf TN~J0408-2418:} This is the z=2.44 source MRC 0406-244 (\cite{mcc96}). The bright object on the POSS is a foreground star to the north-east of the R=22.7 galaxy.

\noindent{\bf TN~J0443-1212:} Using the higher resolution VLA image, we can identify this radio source with a faint object on the POSS.

\noindent{\bf TN~J2106-2405:} This is the z=2.491 source MRC 2104-242 (\cite{mcc96}). The identification is an R=22.7 object, not the star to the north-north-west of the NVSS position.
\begin{figure}
\psfig{file=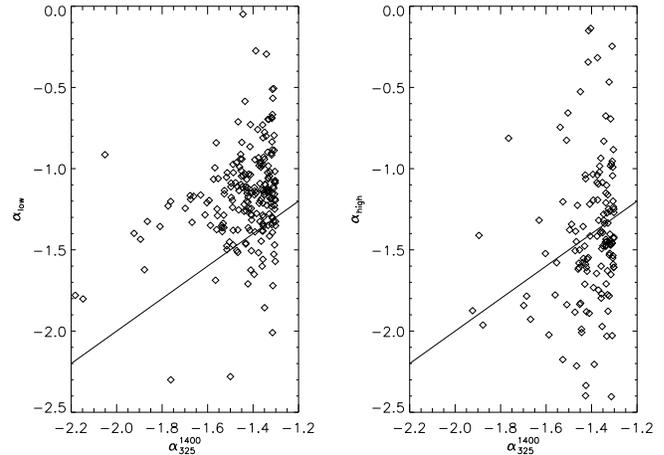,width=9cm,angle=90}
\caption[wnspeccurv]{Radio ``color-color'' plots for the WN sample. The abscissa is the $\alpha_{325}^{1400}$ spectral index used to construct the sample. The ordinates are the low-frequency spectral indices determined from the 8C (38 MHz, \cite{ree90}) or 6C (151 MHz, \cite{hal93}) and the 325~MHz WENSS (left panel), and the high frequency spectral index determined between the 1.4~GHz NVSS and our 4.86~GHz VLA observations (right panel). The line in each panel indicates a straight power law spectrum. Note the unequal number of points on either side of these lines, indicating substantial spectral curvature. \label{wnspeccurv}}
\end{figure}

\subsection{Radio spectra and spectral curvature}

We have used the CATS database at the Special Astronomical Observatory (\cite{ver97}) to search for all published radio measurements of the sources in our samples. In Appendix C, we show the radio spectra for all sources with flux density information for more than two frequencies (the $S_{4860}$ points from our VLA observations are also included). These figures show that most radio spectra have curved spectra, with flatter spectral indices below our selection frequencies, as has been seen in previous USS studies (see \eg\ \cite{rot94}, \cite{blu98}). 

This low frequency flattening and high frequency steepening is obvious in the radio 'color-color diagrams' of the WN sample (Fig. \ref{wnspeccurv}). The median spectral index at low frequencies ($\nu < 325$~MHz) is $-1.16$, while the median $\bar{\alpha}_{325}^{1400} = -1.38$. At higher frequencies ($\nu > 1400$~MHz), the steepening continues to a median $\bar{\alpha}_{1400}^{4850} = -1.44$. Note that the real value of the latter is probably even steeper, as 30\% of the WN sources were not detected in our 4.86~GHz VLA observations, and may therefore have even more steepened high-frequency spectral indices.

\subsection{Radio source properties}
\subsubsection{Radio source structure and angular size}

In Table 5, we give the distribution of the radio structures of the 410 USS sources for which we have good radio-maps. At first sight, all three our samples have basically the same percentage of resolved sources, but the similar value for the MP sample is misleading, as it was observed at much lower resolution.

\begin{table}
\centerline{\bf Table 5: Radio Structure Distribution}
\tiny
\begin{center}
\begin{tabular}{lrrrr}
\\
 & \multicolumn{4}{c}{USS Samples} \\
\cline{2-5} \\
Morphology & WN$\qquad$ & TN$\qquad$ & MP$\qquad$ & Combined$\quad$\\
\hline
Single   & 157 (56$\pm$4\%) & 43 (48$\pm$7\%) & 23 (56$\pm$12\%) & 223 (54$\pm$4\%) \\
Double   &  81 (29$\pm$3\%) & 28 (31$\pm$6\%) & 16 (39$\pm$10\%) & 125 (31$\pm$3\%) \\
Triple   &  22 ( 8$\pm$2\%) &  9 (10$\pm$3\%) &  0 ( 0$\pm$ 0\%) &  31 ( 8$\pm$1\%) \\
Multiple &   2 ( 1$\pm$1\%) &  4 ( 5$\pm$2\%) &  0 ( 0$\pm$ 0\%) &   6 ( 1$\pm$1\%) \\
Diffuse  &  18 ( 6$\pm$2\%) &  5 ( 6$\pm$3\%) &  2 ( 5$\pm$ 3\%) &  25 ( 6$\pm$1\%) \\
\hline
\# Observed & 280 & 89 & 41 & 410 \\
\hline
\end{tabular}
\end{center}
\end{table}

\begin{figure}
\psfig{file=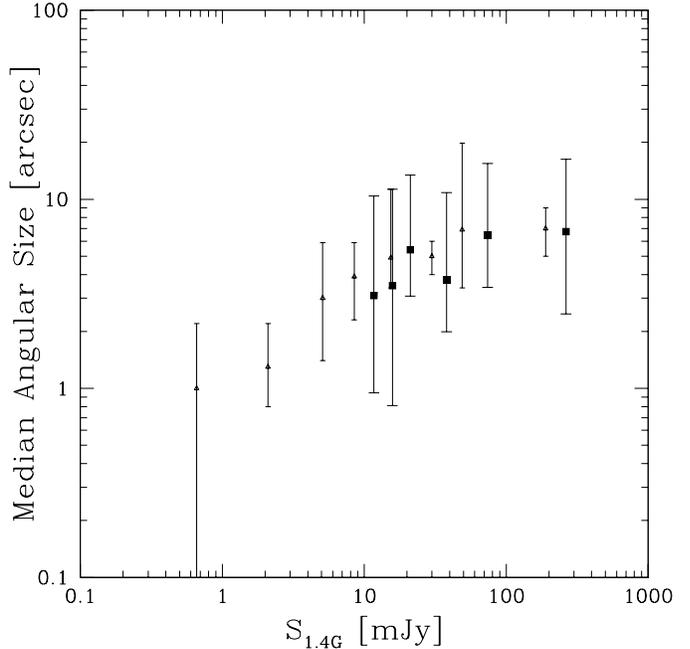,width=9cm}
\caption[slas]{Median angular size for the flux density limited, spectrally unbiased WSRT samples of \cite{oor88} (open triangles), and for our combined USS samples (filled squares). The sources have been binned in equal number bins, and errors represent the 35\% and 65\% levels of the distribution. Note that our USS selection does not affect the value of the median, and that our USS samples also exclude sources that fall below the break at $S_{1400} \lesssim 10$~mJy.  \label{slas}}
\end{figure}

Our results are different from the USS sample of R\"ottgering \etal\ (1994), which contains only 18\% unresolved sources at comparable resolution (1\farcs5). To check if this effect is due to the fainter sources in our sample, we compared our sample with the deep high resolution VLA observations of spectrally unbiased sources (\cite{oor88}; \cite{col85}).
The resolution of our observations is significantly better than the median angular size for $S_{1400} > 1$~mJy sources, allowing us to accurately determine the median angular sizes in our samples.
We find that our USS sources have a constant median angular size of $\sim6$\arcsec\ between 10~mJy and 1~Jy (Fig. \ref{slas}). This is indistinguishable from the results from samples without spectral index selection. It indicates that our USS selection of sources with $\alpha < -1.3$ and $\Theta < 1\arcmin$ does not bias the angular size distribution in the resulting sample.  The 'downturn' in angular sizes that occurs at $\sim$1~mJy is probably due to a different radio source population, which consist of lower redshift sources in spiral galaxies (see \eg, \cite{col85}, \cite{oor87}, \cite{ben93}). By selecting only sources with $S_{1400} > 10$~mJy, we have avoided ``contamination'' of our sample by these foreground sources.

We have searched for further correlations between spectral index or spectral curvatures and angular size or flux density, but found no significant results, except for a trend for more extended sources to have lower than expected 4.86~GHz flux densities, but this effect can be explained by missing flux at large scales in our VLA observations.

\subsection{Identifications}
\subsubsection{POSS}
We have searched for optical identifications of our USS sources on the digitized POSS-I. We used the likelihood ratio identification criterion as described by \eg de Ruiter \etal\ (1977). In short, this criterion compares the probability that a radio and optical source with a certain positional difference are really associated with the probability that this positional difference is due to confusion with a field object (mostly a foreground star), thereby incorporating positional uncertainties in both radio and optical positions. The ratio of these probabilities is expressed as the likelihood ratio $LR$. 
In the calculation, we have assumed a density of POSS objects $\rho = 4 \times 10^{-4}$ \arcsec$^{-2}$, independent of galactic latitude $b$. 
We have adopted a likelihood ratio cutoff \Lh=1.0, slightly lower than the values used by de Ruiter \etal\ (1977) and R\"ottgering \etal\ (1994). We list sources with $LR >$ 1.0 for our USS samples in Tables A.4 to A.6. We have included four WN sources (WN~J0704+6318, WN~J1259+3121, WN~J1628+3932 and WN~J2313+3842), two TN sources (TN~J0510-1838 and TN~J1521+0741) and four MP sources (MP~J0003-3556, MP~J1921-5431, MP~J1943-4030 and MP~J2357-3445) as identifications because both their optical and radio morphologies are diffuse and overlapping, making it impossible to measure a common radio and optical component, while they are very likely to be associated.

\begin{figure}
\psfig{file=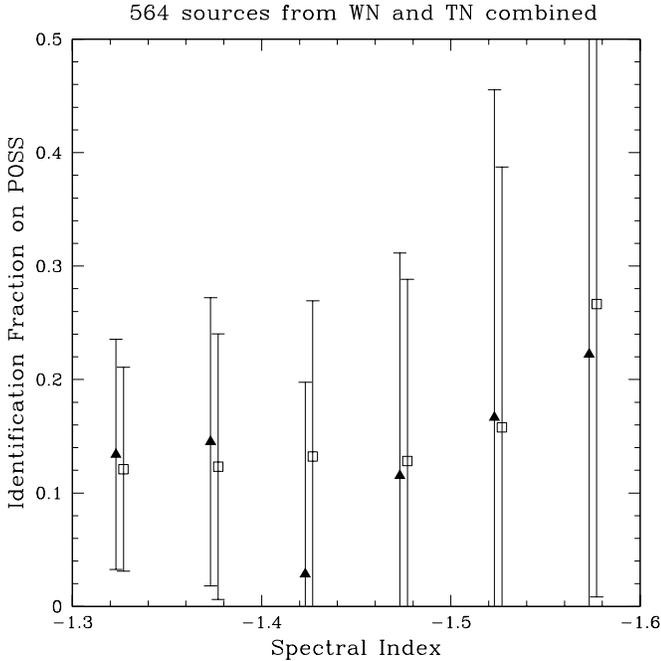,width=9cm}
\caption[idfraction]{Identification fraction on the POSS as a function of spectral index for the combined WN and TN sample. Note the absence of a further decrease in the identification fraction with steepening spectral index. \label{idfraction}}
\end{figure}

Figure \ref{idfraction} shows the identification fraction of USS sources on the POSS ($R \lesssim 20$). Because the distributions for the WN and TN are very similar, we have combined both samples to calculate the identification fraction. Unlike the results for 4C USS (\cite{tie79,blu79}), we do not detect a decrease of the identification fraction with steepening spectral index\footnote{In the Westerbork faint USS (\cite{wie92}) or the USS sample from R\"ottgering \etal\ (1994), there is also a decrease in the identification fraction, even at limiting magnitudes of $R=22.5$ and $R=23.7$, indicating that this trend continues out to fainter magnitudes and radio fluxes}. We interpret the constant $\sim$15\% identification fraction from our sample as a combined population of foreground objects, primarily consisting of clusters (see next section). Our extremely steep spectral index criterion would then selected only radio galaxies too distant to be detected on the POSS ($R \simgt 20.0$).

\subsubsection{Literature}
Using the NASA Extragalactic Database (NED), the SIMBAD database and the W$^3$Browse at the High Energy Astrophysics Science Archive Research Center, we have searched for known optical and X-ray identifications of sources in our samples (see appendices A.7 to A.9). Of the bright optical ($R \lesssim 20$) identifications, only one source is a known as a K0-star, three (TN~J0055$+$2624, TN~J0102$-$2152, and TN~J1521$+$0742) are ``Relic radio galaxies'' (\cite{kom94}, \cite{gio99}), while all others are known galaxy clusters. 

All optical cluster identifications, except MP~J1943-4030, are also detected in the ROSAT All-Sky survey Bright Source Catalogue (RASS-BSC; \cite{vog99}). Conversely, of the 23 X-ray sources, seven are known galaxy clusters, and three known galaxies. The remaining 13 sources are good galaxy cluster candidates because they either show a clear over-density of galaxies on the POSS (eight sources), or they have low X-ray count rates ($<$ 0.02 counts s$^{-1}$), suggesting that these might be more distant galaxy clusters too faint to be detected on the POSS. We conclude that probably $>$3\% of our USS sources are associated with galaxy clusters, and that the combined USS + X-ray selection is an efficient (up to 85\%) selection technique to find galaxy clusters\footnote{In the RASS-BSC, only 14\% of the extra-galactic sources are identified with galaxy clusters (\cite{vog99}).}.

\begin{figure}
\psfig{file=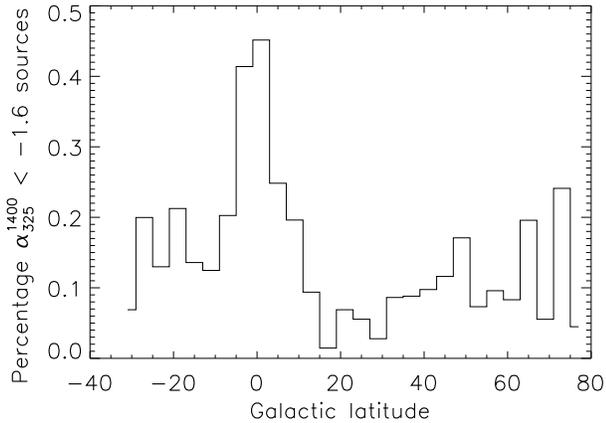,width=9cm}
\caption[pulsargb]{Percentage of $\alpha_{325}^{1400} < -1.60$ radio sources from a WENSS--NVSS correlation as a function of Galactic latitude. Note the clear peak near the Galactic plane, indicating that these $\alpha_{325}^{1400} < -1.60$ objects might well be Galactic pulsars. \label{pulsargb}}
\end{figure}

Three of our USS sources (WN~J2313+4253, TN~J0630-2834 and TN~J1136+1551) are previously known pulsars (\cite{kap98}). It is worth noting that two out of nine sources in our USS samples with $\alpha < -2$ are known pulsars. Because Lorimer \etal\ (1995) \nocite{lor95} found the median spectral index of pulsars to be $\sim -1.6$, we examined the distribution of spectral indices as a function of Galactic latitude. In figure \ref{pulsargb}, we plot the percentage of $\alpha_{325}^{1400} < -1.60$ pulsar candidates as a function of Galactic latitude. The four times higher density near the Galactic plane strongly suggests that the majority of these $\alpha_{325}^{1400} < -1.60$ sources are indeed pulsars, which are confined to our Galaxy. A sample of such $\alpha_{325}^{1400} < -1.60$ sources at $|b|<15$\degr\ would be an efficient pulsar search method.

We also note that no known quasars are present in our sample. Preliminary results from our optical spectroscopy campaign (De Breuck \etal\ 1998b, 2000) indicate that $\sim$10\% of our sample are quasars. We interpret this lack of previously known quasars are a selection bias in quasar samples against USS sources.

At $R \simgt 20$, all five USS sources with known redshift are HzRGs, indicating a selection of sources without detections on the POSS strongly increases our chances of finding HzRGs.

\section{Conclusions}
We have constructed three spatially separated samples of USS sources containing a total of 669 objects. High-resolution radio observations of more than half of these show that the median size is $\sim$6\arcsec, independent of 1.4 GHz flux density, which is consistent with results of similar resolution surveys of samples without spectral index selection. The absence of a downturn in angular size at the lowest fluxes indicates that we do not include significant numbers of spiral galaxies in our sample. A USS sample fainter than ours would therefore include more of these foreground sources, and be less efficient to find HzRGs. 

The identification fraction on the POSS is $\sim$15\%, with no clear dependence on spectral index, indicating that the HzRGs in the sample are all too distant to be detected, and the POSS detections consist of different classes of objects. A correlation of our USS samples with X-ray catalogs showed that at least 85\% of the X-ray identifications seem to be galaxy clusters known from the literature or by inspection of the POSS. We conclude that (1) the majority of the 'non HzRG' USS sources in our sample are clusters, and (2) the combined selection of USS and X-ray sources is an extremely efficient technique to select galaxy clusters.

The above results indicate that up to 85\% of our USS sample might be HzRGs. To identify these objects, we have started an intensive program of R- and K-band imaging on 3--10m class telescopes.
Initial results from optical spectroscopy indicate that 2/3 are indeed $z>2$ radio galaxies (\cite{deb98a}), and K-band imaging of optically undetected ($R > 25$) sources (see \eg, \cite{wvb99a}) has already lead to the discovery of the first radio galaxy at $z>5$ (\cite{wvb99b}). 

\begin{acknowledgements}

We are grateful for the excellent help provided by the staff of the
VLA and ATCA observatories, with special thanks to Chris Carilli and
Greg Taylor (NRAO), and Ray Norris and Kate Brooks (ATNF) for help in
with observation planning and data reduction. We thank Hien ``Napkin''
Tran for his comments on the manuscript. The VLA is a facility of the
National Radio Astronomy Observatory, which is operated by Associated
Universities Inc. under cooperative agreement with the National
Science Foundation. The Australia Telescope is funded by the
Commonwealth of Australia for operation as a National Facility managed
by CSIRO. The authors made use of the database CATS (\cite{ver97}) of
the Special Astrophysical Observatory, the NASA/IPAC Extragalactic
Database (NED) which is operated by the Jet Propulsion Laboratory,
California Institute of Technology, under contract with the National
Aeronautics and Space Administration, and the High Energy Astrophysics
Science Archive Research Center Online Service, provided by the
NASA/Goddard Space Flight Center. Work performed at the Lawrence
Livermore National Laboratory is supported by the DOE under contract
W7405-ENG-48.

\end{acknowledgements}

\appendix
\section{Source Lists}
\begin{table*}
\centerline{\bf Table A.1: WN sample}
\tiny
\begin{tabular}{lrrrrrrrrrrrr}
Name & $S_{325}$ & $S_{1400}$ & S$_{4850}$ & $\alpha_{325}^{1400}$ & $\alpha_{1400}^{4850}$ & Str & LAS & PA & $\alpha_{J2000}$ & $\delta_{J2000}$ & Pos \\
 & mJy & mJy & mJy & & & & \arcsec & $\arcdeg$ & $^h\;\;$$^m\;\;\;\;$$^s\;\;\,$ & \arcdeg$\;\;\;$ \arcmin$\;\;\;$ \arcsec$\;$ & \\
\hline
WN J0000+4654 &   204$\pm$ 9 &   21.1$\pm$ 1.4 & $  2.97\pm0.13$ & $-1.55\pm$0.06 &  $-1.58\pm0.06$ &   D     &  3.6    & 173     & 00 00 28.88 & +46 54 40.7 & V \\
WN J0007+3641 &   447$\pm$18 &   28.8$\pm$ 1.7 & $  2.51\pm0.11$ & $-1.88\pm$0.05 &  $-1.96\pm0.06$ &   S     &  1.7    &  67     & 00 07 02.92 & +36 41 55.9 & V \\
WN J0007+4615 &   144$\pm$ 7 &   16.9$\pm$ 0.5 & \nodata         & $-1.47\pm$0.04 &  \nodata        & \nodata & \nodata & \nodata & 00 07 15.05 & +46 15 40.8 & N \\
WN J0016+4311 &   105$\pm$ 6 &   15.7$\pm$ 1.5 & \nodata         & $-1.30\pm$0.08 &  \nodata        & \nodata & \nodata & \nodata & 00 16 40.63 & +43 11 03.5 & N \\
WN J0029+3439 &   288$\pm$12 &   40.9$\pm$ 2.0 & $ 44.76\pm1.05$ & $-1.34\pm$0.04 &  $ 0.07\pm0.04$ &  DF     & 12.8    &  80     & 00 29 01.61 & +34 39 34.7 & V \\
&&&&&&&&&&&\\
\end{tabular}
\end{table*}

\begin{table*}
\tiny
\begin{tabular}{lrrrrrrrrrrrr}
Name & $S_{325}$ & $S_{1400}$ & S$_{4850}$ & $\alpha_{325}^{1400}$ & $\alpha_{1400}^{4850}$ & Str & LAS & PA & $\alpha_{J2000}$ & $\delta_{J2000}$ & Pos \\
 & mJy & mJy & mJy & & & & \arcsec & $\arcdeg$ & $^h\;\;$$^m\;\;\;\;$$^s\;\;\,$ & \arcdeg$\;\;\;$ \arcmin$\;\;\;$ \arcsec$\;$ & \\
\hline
WN J0911+6306 &   118$\pm$ 6 &   12.9$\pm$ 1.3 & \nodata         & $-1.52\pm$0.08 &  \nodata        & \nodata & \nodata & \nodata & 09 11 34.24 & +63 06 21.7 & N \\
WN J0913+6104 &   297$\pm$12 &   33.8$\pm$ 1.8 & \nodata         & $-1.49\pm$0.05 &  \nodata        & \nodata & \nodata & \nodata & 09 13 05.21 & +61 04 11.7 & N \\
WN J0915+8133 &   127$\pm$ 5 &   18.6$\pm$ 0.6 & $  2.99\pm0.13$ & $-1.32\pm$0.04 &  $-1.47\pm0.04$ &   S     &  2.3    & 127     & 09 15 09.29 & +81 33 34.1 & V \\
WN J0920+5544 &   142$\pm$ 7 &   13.5$\pm$ 1.2 & \nodata         & $-1.61\pm$0.07 &  \nodata        &   D     &  8.9    & 129     & 09 20 46.78 & +55 44 21.8 & F \\
WN J0923+4602 &   107$\pm$ 5 &   12.1$\pm$ 0.5 & \nodata         & $-1.49\pm$0.06 &  \nodata        &   D     & 21.6    & 172     & 09 23 10.44 & +46 02 59.0 & F \\
&&&&&&&&&&&\\
\end{tabular}
\end{table*}

\begin{table*}
\tiny
\begin{tabular}{lrrrrrrrrrrrr}
Name & $S_{325}$ & $S_{1400}$ & S$_{4850}$ & $\alpha_{325}^{1400}$ & $\alpha_{1400}^{4850}$ & Str & LAS & PA & $\alpha_{J2000}$ & $\delta_{J2000}$ & Pos \\
 & mJy & mJy & mJy & & & & \arcsec & $\arcdeg$ & $^h\;\;$$^m\;\;\;\;$$^s\;\;\,$ & \arcdeg$\;\;\;$ \arcmin$\;\;\;$ \arcsec$\;$ & \\
\hline
WN J1327+5341 &   196$\pm$ 9 &   27.4$\pm$ 0.7 & \nodata         & $-1.35\pm$0.04 &  \nodata        &   S     &  0.6    & 162     & 13 27 12.77 & +53 41 20.6 & F \\
WN J1327+5332 &   163$\pm$ 7 &   20.5$\pm$ 0.6 & $  2.70\pm0.17$ & $-1.42\pm$0.04 &  $-1.61\pm0.06$ &   D     &  0.3    &  89     & 13 27 37.86 & +53 32 10.8 & V \\
WN J1329+3046$^*$ &    86$\pm$ 5 &   11.7$\pm$ 1.7 & \nodata         & $-1.37\pm$0.11 &  \nodata        &   S     &  1.6    &   6     & 13 29 31.75 & +30 46 13.6 & F \\
WN J1329+3046$^*$ &    88$\pm$ 5 &   12.1$\pm$ 2.0 & \nodata         & $-1.36\pm$0.13 &  \nodata        &   S     &  3.6    & 129     & 13 29 55.12 & +30 46 51.2 & F \\
WN J1330+6505 &    74$\pm$ 6 &   10.5$\pm$ 0.5 & $  2.34\pm0.12$ & $-1.34\pm$0.06 &  $-1.21\pm0.06$ &   S     &  1.8    &   5     & 13 30 02.77 & +65 04 59.3 & V \\
&&&&&&&&&&&\\
\end{tabular}
\end{table*}

\begin{table*}
\tiny
\begin{tabular}{lrrrrrrrrrrrr}
Name & $S_{325}$ & $S_{1400}$ & S$_{4850}$ & $\alpha_{325}^{1400}$ & $\alpha_{1400}^{4850}$ & Str & LAS & PA & $\alpha_{J2000}$ & $\delta_{J2000}$ & Pos \\
 & mJy & mJy & mJy & & & & \arcsec & $\arcdeg$ & $^h\;\;$$^m\;\;\;\;$$^s\;\;\,$ & \arcdeg$\;\;\;$ \arcmin$\;\;\;$ \arcsec$\;$ & \\
\hline
WN J1717+3828 &   368$\pm$15 &   35.8$\pm$ 2.0 & \nodata         & $-1.60\pm$0.05 &  \nodata        &   M     & 33.6    & 156     & 17 17 16.22 & +38 28 16.5 & F \\
WN J1718+5823 &  1098$\pm$44 &  133.2$\pm$ 2.7 & $ 11.26\pm0.32$ & $-1.44\pm$0.03 &  $-1.98\pm0.03$ &   S     &  0.3    & 107     & 17 18 18.51 & +58 23 21.8 & V \\
WN J1723+5822 &    88$\pm$ 6 &   10.4$\pm$ 0.5 & $  2.27\pm0.11$ & $-1.46\pm$0.06 &  $-1.23\pm0.06$ &   S     &  4.7    & 103     & 17 23 01.98 & +58 22 44.0 & V \\
WN J1723+6844 &   105$\pm$ 6 &   15.7$\pm$ 1.3 & \nodata         & $-1.30\pm$0.07 &  \nodata        & \nodata & \nodata & \nodata & 17 23 05.64 & +68 44 17.7 & N \\
WN J1731+4640 &   200$\pm$ 9 &   15.7$\pm$ 0.5 & \nodata         & $-1.74\pm$0.04 &  \nodata        &   D     & 11.6    & 106     & 17 31 46.10 & +46 40 03.1 & F \\
&&&&&&&&&&&\\
\hline
\end{tabular}

$^{\dag}$ Not a real USS source; see notes

$^*$ See notes

\end{table*}

\begin{table*}
\centerline{\bf Table A.2: TN sample}
\tiny
\begin{tabular}{lrrrrrrrrrrrr}
Name & $S_{365}$ & $S_{1400}$ & S$_{4850}$ & $\alpha_{365}^{1400}$ & $\alpha_{1400}^{4850}$ & Str & LAS & PA & $\alpha_{J2000}$ & $\delta_{J2000}$ & Pos \\
 & mJy & mJy & mJy & & & & \arcsec & $\arcdeg$ & $^h\;\;$$^m\;\;\;\;$$^s\;\;\,$ & \arcdeg$\;\;\;$ \arcmin$\;\;\;$ \arcsec$\;$ & \\
\hline
TN J0008-0912 &   638$\pm$ 64 &  100.2$\pm$ 4.4 & \nodata      & $-1.38\pm$0.08 &  \nodata        & \nodata & \nodata  & \nodata & 00 08 52.43 & $-$09 12 01.4 & N \\
TN J0018+1904 &   498$\pm$ 30 &   84.5$\pm$ 3.8 & \nodata      & $-1.32\pm$0.06 &  \nodata        & \nodata & \nodata  & \nodata & 00 18 07.35 & $+$19 04 50.0 & N \\
TN J0026+1501 &   241$\pm$ 26 &   39.4$\pm$ 2.0 & \nodata      & $-1.35\pm$0.09 &  \nodata        & \nodata & \nodata  & \nodata & 00 26 03.31 & $+$15 01 29.2 & N \\
TN J0027+0059 &   735$\pm$ 55 &   98.0$\pm$ 4.4 & \nodata      & $-1.50\pm$0.07 &  \nodata        & \nodata & \nodata  & \nodata & 00 27 39.36 & $+$00 59 53.6 & N \\
TN J0033-2731 &   512$\pm$ 41 &   66.1$\pm$ 3.0 & \nodata      & $-1.52\pm$0.07 &  \nodata        & \nodata & \nodata  & \nodata & 00 33 24.28 & $-$27 31 48.1 & N \\
&&&&&&&&&&&\\
\end{tabular}
\end{table*}
\begin{table*}
\tiny
\begin{tabular}{lrrrrrrrrrrrr}
Name & $S_{365}$ & $S_{1400}$ & S$_{4850}$ & $\alpha_{365}^{1400}$ & $\alpha_{1400}^{4850}$ & Str & LAS & PA & $\alpha_{J2000}$ & $\delta_{J2000}$ & Pos \\
 & mJy & mJy & mJy & & & & \arcsec & $\arcdeg$ & $^h\;\;$$^m\;\;\;\;$$^s\;\;\,$ & \arcdeg$\;\;\;$ \arcmin$\;\;\;$ \arcsec$\;$ & \\
\hline
TN J0856-1510 &   469$\pm$ 68 &   70.0$\pm$ 3.2 & $ 3.7\pm0.3$ & $-1.41\pm$0.11 &  $-2.01\pm0.07$ &   S     &  1.0     &  14     & 08 56 12.44 & $-$15 10 35.7 & V \\
TN J0902+1809 &   491$\pm$ 37 &   55.5$\pm$ 2.6 & \nodata      & $-1.62\pm$0.07 &  \nodata        & \nodata & \nodata  & \nodata & 09 02 00.00 & $+$18 09 03.4 & N \\
TN J0906-0701 &   939$\pm$101 &  142.5$\pm$ 5.8 & \nodata      & $-1.40\pm$0.09 &  \nodata        & \nodata & \nodata  & \nodata & 09 06 40.23 & $-$07 01 12.4 & N \\
TN J0910-2228 &   465$\pm$ 32 &   55.3$\pm$ 2.6 & $ 4.0\pm0.3$ & $-1.58\pm$0.06 &  $-2.11\pm0.07$ &   S     &  2.0     &   0     & 09 10 34.15 & $-$22 28 47.4 & V \\
TN J0914+1119 &   343$\pm$ 53 &   49.5$\pm$ 1.1 & \nodata      & $-1.44\pm$0.12 &  \nodata        & \nodata & \nodata  & \nodata & 09 14 30.81 & $+$11 19 08.0 & N \\
&&&&&&&&&&&\\
\end{tabular}
\end{table*}
\begin{table*}
\tiny
\begin{tabular}{lrrrrrrrrrrrr}
Name & $S_{365}$ & $S_{1400}$ & S$_{4850}$ & $\alpha_{365}^{1400}$ & $\alpha_{1400}^{4850}$ & Str & LAS & PA & $\alpha_{J2000}$ & $\delta_{J2000}$ & Pos \\
 & mJy & mJy & mJy & & & & \arcsec & $\arcdeg$ & $^h\;\;$$^m\;\;\;\;$$^s\;\;\,$ & \arcdeg$\;\;\;$ \arcmin$\;\;\;$ \arcsec$\;$ & \\
\hline
TN J1439-3226 &   622$\pm$ 65 &   90.7$\pm$ 1.9 & \nodata      & $-1.43\pm$0.08 &  \nodata        & \nodata & \nodata  & \nodata & 14 39 02.87 & $-$32 26 46.6 & N \\
TN J1449+1440 &   464$\pm$ 50 &   80.3$\pm$ 3.7 & \nodata      & $-1.30\pm$0.09 &  \nodata        & \nodata & \nodata  & \nodata & 14 49 02.50 & $+$14 40 42.2 & N \\
TN J1451+2351 &   595$\pm$ 44 &   93.2$\pm$ 4.2 & \nodata      & $-1.38\pm$0.07 &  \nodata        &   D     & 12.6     & 109     & 14 51 10.71 & $+$23 51 38.8 & F \\
TN J1452-1113 &   344$\pm$ 55 &   53.9$\pm$ 2.6 & \nodata      & $-1.38\pm$0.13 &  \nodata        &  S      & $<$ 7.5  &   2     & 14 52 04.88 & $-$11 13 35.7 & A \\
TN J1452+2013 &   291$\pm$ 40 &   43.4$\pm$ 1.0 & \nodata      & $-1.42\pm$0.11 &  \nodata        & \nodata & \nodata  & \nodata & 14 52 22.93 & $+$20 13 09.8 & N \\
&&&&&&&&&&&\\
\end{tabular}
\end{table*}
\begin{table*}
\tiny
\begin{tabular}{lrrrrrrrrrrrr}
Name & $S_{365}$ & $S_{1400}$ & S$_{4850}$ & $\alpha_{365}^{1400}$ & $\alpha_{1400}^{4850}$ & Str & LAS & PA & $\alpha_{J2000}$ & $\delta_{J2000}$ & Pos \\
 & mJy & mJy & mJy & & & & \arcsec & $\arcdeg$ & $^h\;\;$$^m\;\;\;\;$$^s\;\;\,$ & \arcdeg$\;\;\;$ \arcmin$\;\;\;$ \arcsec$\;$ & \\
\hline
TN J2301+0555 &   484$\pm$ 62 &   74.1$\pm$ 3.3 & \nodata      & $-1.40\pm$0.10 &  \nodata        & \nodata & \nodata  & \nodata & 23 01 18.11 & $+$05 55 01.0 & N \\
TN J2303-0251 &   405$\pm$ 56 &   66.5$\pm$ 3.1 & \nodata      & $-1.34\pm$0.11 &  \nodata        & \nodata & \nodata  & \nodata & 23 03 55.67 & $-$02 51 08.6 & N \\
TN J2310+1358 &  1497$\pm$102 &  251.9$\pm$ 5.2 & \nodata      & $-1.33\pm$0.05 &  \nodata        & \nodata & \nodata  & \nodata & 23 10 52.32 & $+$13 58 20.5 & N \\
TN J2312-0040 &   293$\pm$ 25 &   45.5$\pm$ 1.0 & \nodata      & $-1.39\pm$0.07 &  \nodata        & \nodata & \nodata  & \nodata & 23 12 46.90 & $-$00 40 53.0 & N \\
TN J2312-0208 &  1295$\pm$ 80 &  194.8$\pm$ 8.6 & \nodata      & $-1.41\pm$0.06 &  \nodata        & \nodata & \nodata  & \nodata & 23 12 55.71 & $-$02 08 39.9 & N \\
&&&&&&&&&&&\\
\hline
\end{tabular}

$^{\dag}$ Not a real USS source; see notes

$^*$ See notes
\vspace{1cm}
\end{table*}
\begin{table*}
\centerline{\bf Table A.3: MP sample}
\tiny
\begin{tabular}{rrrrrrrrrrrrr}
Name & $S_{408}$ & $S_{1420}$ & $S_{4850}$ & $\alpha_{408}^{1420}$ & $\alpha_{408}^{4850}$ & Str & LAS & PA & $\alpha_{J2000}$ & $\delta_{J2000}$ & Pos\\
 & mJy & mJy & mJy & & & & \arcsec & $\arcdeg$ & $^h\;\;$$^m\;\;\;\;$$^s\;\;\,$ & \arcdeg$\;\;\;$ \arcmin$\;\;\;$\arcsec$\;$ & \\
\hline
MP J0003-3556 &   2440$\pm$ 80 & \nodata      & 103.0$\pm$13.0 & \nodata        & $-1.28\pm$0.05 & \nodata & \nodata & \nodata & 00 03 13.11 & $-$35 56 33.8 & M \\
MP J0028-5523 &   1910$\pm$ 80 & $ 389\pm 38$ &  86.0$\pm$ 9.0 & $-1.28\pm0.10$ & $-1.25\pm$0.05 &  S      &   3.7   &  57     & 00 28 53.76 & $-$55 23 28.5 & A \\
MP J0100-6403 &   1410$\pm$150 & $ 247\pm 24$ &  56.0$\pm$ 7.0 & $-1.40\pm0.11$ & $-1.30\pm$0.07 &  S      &   4.2   & 100     & 01 00 32.42 & $-$64 03 13.6 & A \\
MP J0103-3225 &   2290$\pm$ 80 & \nodata      &  93.0$\pm$13.0 & \nodata        & $-1.29\pm$0.06 & \nodata & \nodata & \nodata & 01 03 17.10 & $-$32 25 53.3 & M \\
MP J0114-3302 &   1590$\pm$ 90 & $ 396\pm 39$ &  76.0$\pm$12.0 & $-1.11\pm0.14$ & $-1.23\pm$0.07 &  D      &  22.0   & 119     & 01 14 37.06 & $-$33 02 10.2 & A \\
&&&&&&&&&&&\\
\hline
\end{tabular}
\end{table*}

\begin{table}
\footnotesize
\centerline{\bf Table A.4: WN POSS identifications}
\begin{tabular}{rrrrr}
Name & $S_{1400}$ & $\alpha_{325}^{1400}$ & LAS & LR$^a$ \\
 & mJy & & \arcsec & \\
\hline
WN J0029$+$3439 &   40.9 & $-$1.34 & 12.8 &   4.2 \\
WN J0121$+$4305 &   50.8 & $-$1.39 & \nodata &   7.1 \\
WN J0315$+$3757 &   13.3 & $-$1.35 &  1.7 &  14.8 \\
WN J0559$+$6926 &   12.9 & $-$1.31 &  0.3 &   4.6 \\
WN J0641$+$4325 &   15.6 & $-$1.57 & \nodata & 195.6 \\
WN J0725$+$4123 &   35.5 & $-$1.39 &  8.3 & 183.0 \\
\hline
\end{tabular}

$^a$ Likelihood Ratio, see text
\vspace{1cm}
\end{table}

\begin{table}
\footnotesize
\centerline{\bf Table A.5: TN POSS identifications}
\begin{tabular}{rrrrr}
Name & $S_{1400}$ & $\alpha_{365}^{1400}$ & LAS & LR$^a$ \\
 & mJy & & \arcsec & \\
\hline
TN J0244$+$0327 &  84.8 & $-$1.38 & \nodata &   6.9 \\
TN J0245$+$2700 &  54.0 & $-$1.47 & \nodata &  30.5 \\
TN J0250$+$0130 &  39.1 & $-$1.45 & \nodata &  14.7 \\
TN J0256$-$2717 & 177.9 & $-$1.32 & \nodata & 126.6 \\
TN J0301$+$0155 & 402.7 & $-$1.70 & \nodata &  21.3 \\
TN J0433$+$0717 &  33.1 & $-$1.44 & \nodata &  29.9 \\
\hline
\end{tabular}

$^a$ Likelihood Ratio, see text
\vspace{1cm}
\end{table}

\begin{table}
\footnotesize
\centerline{\bf Table A.6: MP UKST identifications}
\begin{tabular}{rrrrr}
Name & $S_{408}$ & $\alpha_{408}^{4850}$ & LAS & LR$^a$ \\
 & mJy & & \arcsec & \\
\hline
MP J0003$-$3556 & 2440 & $-$1.28 & \nodata &   0.0 \\
MP J0103$-$3225 & 2290 & $-$1.29 & \nodata &  88.1 \\
MP J0618$-$7340 &  750 & $-$1.37 & \nodata &   7.5 \\
MP J1033$-$3418 & 5590 & $-$1.23 & \nodata &  81.7 \\
MP J1250$-$4026 & 1350 & $-$1.24 &     7.4 & 131.0 \\
MP J1921$-$5431 & 4590 & $-$1.71 & \nodata &   0.0 \\
MP J1943$-$4030 & 4400 & $-$1.36 & \nodata &   0.0 \\
MP J2313$-$4243 & 1390 & $-$1.28 & \nodata &  95.5 \\
MP J2357$-$3445 & 8700 & $-$1.74 & \nodata &   0.0 \\
\hline
\end{tabular}

$^a$ Likelihood Ratio, see text
\vspace{1cm}
\end{table}

\begin{table*}
\centerline{\bf Table A.7: WN identifications from the literature}
\begin{tabular}{rrrrrr}
Name & $z$ & $R$[mag] & $F_X^a$ & Identification & Reference \\
\hline
WN J0633$+$4653 & \nodata &    23.6 & \nodata &         4C +46.12 & \cite{cha96a} \\
WN J0648$+$4309 & \nodata &    23.0 & \nodata &       B3 0644+432 & \cite{wie92} \\
WN J0658$+$4444 & \nodata &    23.4 & \nodata &       B3 0654+448 & \cite{wie92} \\
WN J0717$+$4611 &  1.462  &    21.6 & \nodata &       B3 0714+462 & \cite{deb98b} \\
WN J0725$+$4123 & \nodata & \nodata &  0.1137 &  1RXS J072600.0+4 & \cite{vog99} \\
WN J0923$+$4602 & \nodata & \nodata &  0.0242 &  1WGA J0923.1+460 & \cite{whi94}\\
WN J0952$+$5153 &  0.214  & \nodata &  0.2413 &  ZwCl 0949.6+5207 & \cite{vog99} \\
\hline
\end{tabular}

$^a$F$_X$ is the number of X-ray counts s$^{-1}$ as listed in the cited catalogs.

\vspace{1cm}
\end{table*}

\begin{table*}
\centerline{\bf Table A.8: TN identifications from the literature}
\begin{tabular}{rrrrrr}
Name & $z$ & $R$[mag] & $F_X^a$ & Identification & Reference \\
\hline
TN J0055$+$2624 & 0.1971  & \nodata & \nodata &        ABELL  85 & \cite{gio99} \\
TN J0102$-$2152 & 0.0604  & \nodata & \nodata &        ABELL 133 & \cite{kom94} \\
TN J0245$+$2700 & \nodata & \nodata &  0.0782 & 1RXS J024521.0+2 & \cite{bad98} \\
TN J0256$-$2717 &   0.480 &    19.2 & \nodata &      MRC 0254-27 & \cite{mcc96} \\
TN J0301$+$0155 &  0.170  & \nodata &  0.2053 & ZwCl 0258.9+0142 & \cite{cra95} \\
TN J0408$-$2418 &  2.440  &    22.7 & \nodata &     MRC 0406-244 & \cite{mcc96} \\
TN J0510$-$1838 & \nodata &    19.5 & 0.00314 & 1RXS J051032.4-1 & \cite{vog99} \\
TN J0630$-$2834 & \nodata & \nodata & 0.00314 & 1WGA J0630.8-283 & \cite{whi94} \\
TN J0729$+$2436 & \nodata & \nodata &  0.1752 & 1RXS J072927.4+2 & \cite{vog99} \\
TN J0936$-$2243 &  1.339  &    23.0 & \nodata &           3C 222 & \cite{hec94} \\
TN J0958$-$1103 &  0.153  & \nodata &    1.64 &       ABELL 0907 & \cite{mcc96} \\
TN J1521$+$0742 &  0.045  &    14.5 &    1.64 &         NGC 5920 & \cite{kom94} \\
TN J2106$-$2405 &  2.491  &    22.7 & \nodata &     MRC 2104-242 & \cite{mcc96} \\
TN J2239$-$0429 & \nodata &    18.8 & \nodata &     4C -04.85    & \cite{rot95} \\
TN J2320$+$1222 & \nodata &    22.9 & \nodata &     MRC 2317+120 & \cite{rot95} \\
\hline
\end{tabular}

$^a$F$_X$ is the number of X-ray counts s$^{-1}$ as listed in the cited catalogs.

\vspace{1cm}
\end{table*}

\begin{table*}
\centerline{\bf Table A.9: MP identifications from the literature}
\begin{tabular}{rrrrrr}
Name & $z$ & $R$[mag] & $F_X^a$ & Identification & Reference \\
\hline
MP J0003$-$3556 &  0.0497 &    13.2 &   0.5206 &       ABELL 2717 & \cite{col95} \\
MP J0103$-$3225 & \nodata &    19.0 &          & IRAS F01009-3241 & \cite{con95} \\
MP J1250$-$4026 & \nodata & \nodata & 0.000595 & 1RXP J125006-402 & \cite{vog99} \\
MP J1943$-$4030 & \nodata &    17.3 &          &       ABELL 3646 & \cite{abe89} \\
MP J2313$-$4243 &  0.0564 &    14.6 &    2.024 &      ABELL S1101 & \cite{sto91} \\
MP J2357$-$3445 &  0.0490 &    13.7 &     2.27 &       ABELL 4059 & \cite{pos95} \\
\hline
\end{tabular}

$^a$F$_X$ is the number of X-ray counts s$^{-1}$ as listed in the cited catalogs.

\end{table*}

\clearpage

\section{Radio Maps}
\subsection{VLA Maps of the WN Sample}

\small{VLA maps of the WN sample. The contour scheme is a geometric progression in $\sqrt 2$, which implies a factor 2 change in surface brightness every 2 contours. The first contour level, indicated above each plot, is at $3\sigma_{rms}$, where $\sigma_{rms}$ is the rms noise determined around the sources. The restoring beams are indicated in the lower left corner of the plots. Two maps are given for each source, one showing a 6\arcmin\ field of view to show possible related components, and a smaller blow-up of the source to show its morphology. The open cross indicates the NVSS position. Sources identified on the POSS have been marked in the top right corner. \label{wnvla}}

\subsection{FIRST Maps of the WN Sample}

\small{FIRST maps of the WN sample.Contours are as in section \ref{wnvla} \label{wnfirst}}

\subsection{VLA Maps of the TN Sample}

\small{VLA maps of the TN sample. Contours are as in section \ref{wnvla} \label{tnvla}}
   
\subsection{FIRST Maps of the TN Sample}

\small{FIRST maps of the TN sample. Contours are as in section \ref{wnvla}. \label{tnfirst}}

\subsection{ATCA Maps of the TN Sample}

\small{ATCA maps of the TN sample. Contours are as in section \ref{wnvla}. \label{tnatca}}

\subsection{ATCA Maps of the MP Sample}

\small{ATCA maps of the MP sample. Contours are as in section \ref{wnvla}. \label{mpatca}}

\section{Radio Spectra}
\subsection{Radio Spectra for the WN Sample}

\small{Radio spectra of the WN sample using data from the literature. The two connected flux points indicate the spectral index used to select the source in the USS sample. Note the steeper spectra with higher frequency in most objects. \label{wnradiospectra}}

\subsection{Radio Spectra for the TN Sample}

\small{Radio spectra of the TN sample using data from the literature. The two connected flux points indicate the spectral index used to select the source in the USS sample. Note the steeper spectra with higher frequency in most objects. \label{tnradiospectra}}

\subsection{Radio Spectra for the MP Sample}

\small{Radio spectra of the MP sample using data from the literature and our ATCA observations (diamonds). The two connected flux points indicate the spectral index used to select the source in the USS sample. Note the steeper spectra with higher frequency in most objects. \label{mpradiospectra}}

\section{POSS Finding Charts}
\subsection{POSS Finding Charts for the WN Sample}

\small{POSS finding charts of the WN sample. The open cross indicates the NVSS position.}

\subsection{POSS Finding Charts for the TN Sample}

\small{POSS finding charts of the TN sample. The open cross indicates the NVSS position.}

\subsection{POSS Finding Charts for the MP Sample}

\small{POSS finding charts of the MP sample. The open cross indicates the MRC position.}

\end{document}